\documentclass[
 aip,
 jap,%
 reprint,%
]{revtex4-1}
   \usepackage[pdftex]{graphicx}

\usepackage{comment}

\usepackage{amsmath}
\usepackage{tikz}
\usepackage{tikz-3dplot}

\begin{document}
\title{\LARGE Analysis of a Wireless Power Transfer System Based on the Interaction of Very High Permittivity Dielectric Resonators with a Contained Aqueous Solution}


\author{Sameh. Y. Elnaggar}
\email{samehelnaggar@gmail.com}
\affiliation{Department of Electrical and Computer Enginering, Royal Military College of Canada, Kingston, ON, Canada.}
\author{Chinmoy Saha}
\email{csaha@ieee.org}
\affiliation{Indian Institute of Space Science and Technology,Thiruvananthapuram, Kerala, India.}
\author{Yahia. M. M. Antar}
\email{antar-y@rmc.ca}
\affiliation{Department of Electrical and Computer Enginering, Royal Military College of Canada, Kingston, ON, Canada.}

\date{\today}






\begin{abstract}
An \emph{ab-initio} analysis based on coupled mode theory (CMT) is applied to describe the interaction dynamics of high dielectric resonators (DRs) with its containing aqueous solution. We prove that the coupling mechanism is reciprocal. Such property is exploited to find closed form and accurate expressions of the coupling coefficient $\kappa$, the main factor characterizing the system performance. Based on such expressions, it is shown that, for wireless power transfer (WPT) applications, up sizing the DRs relaxes the need of using ultra high $\epsilon_r$ materials. The nature of interaction is captured by the coupling matrix, which shows that the behaviour of the system is identical to the ones studied extensively in the literature when the contained aqueous solution is replaced by an enclosing cavity. It follows, as in a typical three coupled resonators setting, that when two DRs are inserted in the aqueous solution a non-bonding mode emerges; hence enabling efficient wireless power transfer (WPT) via the opening of an electromagnetic induced transparency like window. Due to the inevitable situations where the DRs are not symmetrically placed inside the solution, the general eigenvalue problem, with asymmetrically placed DR inserts is solved and the eigenvectors representing the coupled modes are depicted by vectors in a 3D mathematical space where the uncoupled modes represent its basis. Moreover, the strong coupling between the DR inserts and the aqueous medium allows the proposed WPT system to tolerate intrinsic aqueous solution losses and the presence of extraneous objects. Additionally, the value of the load at maximum efficiency is indpendent of the aqueous solution loss tangent, thus tolerating the variation in the medium salinity. The Proposed EIT like scheme can find applications for mid/short range power transfer in/through swimming pools, chemical reactors, fish tanks, etc.

\end{abstract}
\maketitle

%

\section{Introduction}
Wireless power transfer via the use of resonant coupling is basically non-radiative electromagnetic transmission with great potential for diversified applications such as the charging of  electronic gadgets, electric vehicles and powering implants in the human body \cite{hui2013planar,shin2014design,xue2013high}. Wireless transfer of power in a fully enclosed environment, reported in Refs. \onlinecite{xue2013high,Chabalko2017,Sastani2017, mei2017cavity}, has the potential of efficiently charging 3D distributed wireless sensors and other devices in a contained environment such as rooms and satellites. Since the revival of  WPT via inductive resonant coupling, capacitively loaded coils have been widely used in near field WPT schemes,  yielding moderate transfer efficiencies in the sub-wavelength regime \cite{Kurs2007,Karalis2008}. Quite recently, the resonant coupling between high $Q$ DR modes was exploited to demonstrate the feasibility of using DRs as alternatives to capacitively loaded coils \cite{Song2016,Song2016Collosal}. High $Q$ values of the DRs results in an efficient coupling of power. Moreover, the inclusion of an additional resonator (hereafter, relay or mediator) that interacts with both the source and receiver DRs can stretch the transfer distances. The interaction between the source, receiver and relay resonators opens an EIT like channel between the source and receiver through the creation of a non-bonding mode \cite{Elnaggar2017JAP}. The creation of the non-bonding (\emph{dark}) mode was explored in an early work in the context of inductively resonant coupled coils, where the coupling coefficient $\kappa$ is moderate \cite{Hamam2009}. In this case, the frequencies of coupled modes are very close and the excitation of the non-bonding mode only is challenging and may need tedious mechanical arrangements. Nevertheless, it has been shown that, in general, the inclusion of the mediator does improve the efficiency and extends the transfer distance, even with no specific arrangements \cite{Zhang2012}.

In the context of Electron Spin Resonance Spectroscopy, it was shown that the ${TE}_{01\delta}$ of a DR inserted in  a cylindrical cavity strongly couples with the cavity ${TE}_{011}$ mode via the overlap of the electric field \cite{Elnaggar2014coupled,Elnaggar2014Quality}. Additionally, placing two DRs inside the cavity generates bonding, non-bonding and anti-bonding modes; hence generating an EIT like channel between the two DRs \cite{AMR2017}. Combining both desirable properties: the strong coupling and the generation of EIT like channel, a WPT scheme was reported in Refs. \onlinecite{Elnaggar2017JAP,samehwptmeas}. The scheme relies on the interaction between two DRs with a split cavity resonator (SCR) that acts as the mediator. The unique properties of the system allow it to be tolerant to losses, material imperfections and frequency offsets between different components \cite{Elnaggar2017JAP, wptc1}.

Recently, it was shown via basic theory and finite element simulations that strong coupling combined with the presence of an EIT like channel do exist in a system comprised of a contained high $\epsilon_r$ aqueous solution and two very high $\epsilon_r\approx 2300$ DR inserts \cite{wptc2}. The DR inserts (hereafter denoted by DR1 and DR3) form the source and receiver, respectively, and the aqueous solution (DR2) acts as the mediator. The high $\epsilon_r$ of the medium imposes a challenge on the permissible DRs to be used. Ultra high $\epsilon_r$ materials can be realized through the use of various composite materials, thanks to pioneering research by various material research groups, that exhibit very high dielectric constant values at microwave frequency range. For example, i) Strontium titanate ($\textnormal{SrTiO}_3$), ii) Barium titanate, iii) Calcium copper titanate (CCTO) and iv) Barium strontium titanate are reported to have  dielectric constant of  20,000, 18000, 10000 and 3000 respectively in lower microwave frequency band at low and moderate temperature \cite{robertsdielectric,yimnature,singh,shende2001strontium}. However, acquiring such materials for experimentation is quite challenging. Hence, the necessity to understand the trade off between the DR $\epsilon_r$ and its dimensions becomes clear.

 In the current article, the main features of the scheme is explained  using an analytical model based on CMT \cite{Elnaggar2015ECMT}. The model is verified by finite element simulations. Closed form expressions of $\kappa$ is derived to reveal its dependency on $\epsilon_r$ and dimensions, and to explain the physical interactions between the resonators. Furthermore realistic scenarios, such as misalignments and the presence of extraneous objects are considered and their impact on the performance of the proposed WPT system is thoroughly investigated. Additionally a geometrical representation of the eigenvectors is proposed to develop a more tangible understanding of the modal behaviour and how they depend on $\kappa$.

Section II focuses on the coupled modes due to the interaction of one DR insert (DR1 or DR3) with the aqueous solution (DR2). In this section closed form expressions for $\kappa$ are derived and verified. The reciprocal nature of interaction is presented and the effect of the medium losses on DR1 intrinsic $Q$ factor is discussed. Section III explores the properties of the modes when all components are present (DR1, DR2 and DR3). The eigenvalue problem is solved in the general case and a geometric representation of the field of the coupled modes is proposed to better highlight their properties. The influence of the non-bonding mode on the transfer efficiency is briefly explained. Finally, the effects of offsets and the presence of extraneous objects are demonstrated via finite element simulations. The conclusion follows in Section IV.
\begin{figure}
\centering
\includegraphics[width=3.0in]{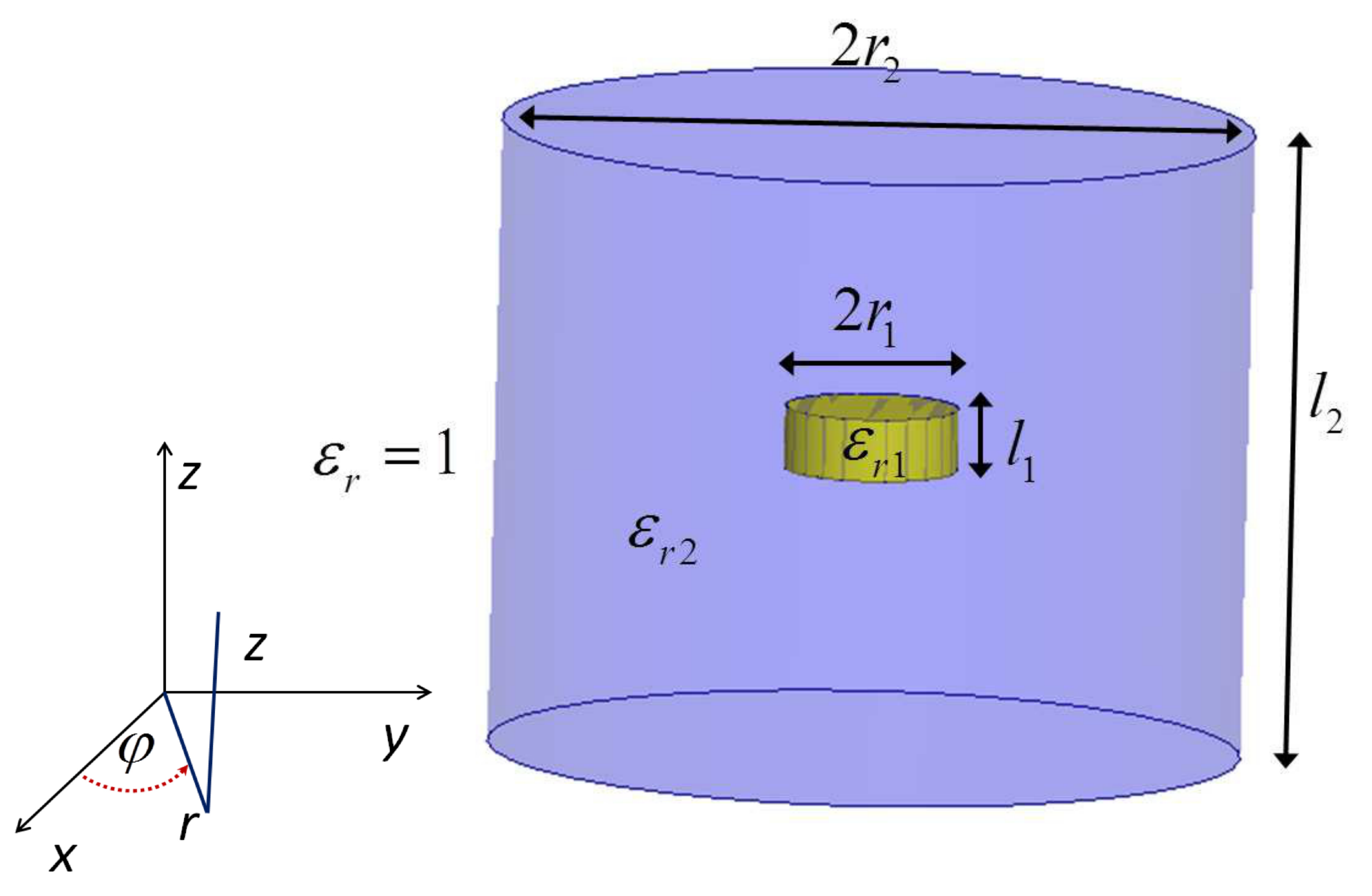}
\caption{A contained aqueous medium, with parameters identified by the subscript '2'. A DR insert (DR1) is concentrically placed within the contained medium.}
\label{fig:DRinDR}
\end{figure}

\section{Interaction of a DR and the contained aqueous medium}
In Ref. \onlinecite{wptc2} the coupling coefficient $\kappa$ between a contained aqueous medium and a DR insert was determined based on physical arguments only. In the current article, however, a systematic route will be taken to describe the interaction between the different components. Not only does the systematic treatment base the findings on a rigorous foundation, it also highlights the assumptions and approximations used along the way. The system configuration is shown in Fig. \ref{fig:DRinDR}. Subsection II-A presents the basic steps necessary to obtain the eigenvalue problem (EVP). In subsection II-B, the EVP is solved and expressions for $\kappa$ are derived.

\subsection{Eigenvalue Problem (EVP)}
In general, rigorous solutions of the eigenmodes of the system in Fig. \ref{fig:DRinDR} can be determined after one solves the Helmholtz's equation
\begin{equation}
\label{eq:Helmholtz}
\nabla\times\nabla\times\mathbf{E}(\mathbf{r})=\frac{\omega^2}{c^2}\epsilon_r(\mathbf{r})\mathbf{E}(\mathbf{r})
\end{equation}
over the entire space after taking into account the conditions dictated by the boundaries. In (\ref{eq:Helmholtz}) $\epsilon_r(\mathbf{r})$ can be written as\cite{wptc2}
\begin{equation}
\label{eq:epsilon}
\epsilon_r(\mathbf{r})=
\begin{cases}\epsilon_{r1} & r\leq r_1 \textnormal{ and } |z- z_0|<l_1/2\\
\epsilon_{r2} &r\leq r_2 \textnormal{ and } |z|<l_2/2\\
1 & \textnormal{otherwise}
\end{cases},
\end{equation}
where $z_0$ is the offset of the DR center with respect to the enclosed medium center. Equation (\ref{eq:Helmholtz}) is in a generalized eigenvalue problem form. In principle, its solutions represent the eigenfrequencies and the corresponding eigenfields of the complete system. In spite of being the most rigorous approach, the procedure does not convey much information about how the characteristics of the system sub-components (here DR1 and DR2) combine. The solutions are usually obtained via the discretization of the $\nabla\times\nabla$ operator and fields over space, resulting in a \emph{discretized} version of (\ref{eq:Helmholtz}), where the frequencies (or more precisely the square of frequencies) represent the eigenvalues and the fields represent the corresponding eigenvectors. On the other hand, CMT permits the projection of the \emph{unknown} total fields $\mathbf{E}$ and $\mathbf{H}$ onto the modes of the sub-components \cite{Elnaggar2015ECMT,wptc2}. As one is interested in a narrow band response, CMT reduces the solution space to the linear superposition of a handful number of modes with frequencies in the vicinity of the band of interest and that exhibit field profiles that have a net overlap in space and match the excitation profile. 

The CMT main premise is the assumption that the fields of the coupled (combined) system are the linear superposition of the fields of its individual sub-components \cite{Elnaggar2015ECMT}
\begin{equation}
\label{eq:expansionE}
\mathbf{E}=a_1\mathbf{E}_1+a_2\mathbf{E}_2
\end{equation}
and
\begin{equation}
\label{eq:expansionH}
\mathbf{H}=b_1\mathbf{H}_1+b_2\mathbf{H}_2.
\end{equation}
In (\ref{eq:expansionE}) and (\ref{eq:expansionH}), the expansion is limited to two modes only: the $TE_{01\delta}$ mode of each resonator. In general, any DR mode satisfies the source-free Maxwell's equations
\begin{equation}
\nabla \times \mathbf{E}_i=-j\omega_0\mu_0\mathbf{H}_i
\end{equation}
and
\begin{equation}
\nabla\times\mathbf{H}_i=j\omega_0\epsilon_i(\mathbf{r})\mathbf{E}_i,
\end{equation}
where $\omega_0$ is the resonant frequency of the two modes and $\epsilon_i$ is function of space; it is equal to $\epsilon_{ri}\epsilon_0$ inside the $i^\textnormal{th}$ DR materials and $\epsilon_0$ everywhere else. The phase of the fields are chosen such that the fields of the different modes are coherent ($\mathbf{E}$ fields of both modes are in phase and are $90^\circ$ out of phase with the $\mathbf{H}$ fields). The coupled fields $\mathbf{E}$ and $\mathbf{H}$ also satisfy Maxwell's equations
\begin{equation}
\nabla\times\mathbf{E}=-j\omega\mu_0\mathbf{H}
\end{equation}
and
\begin{equation}
\nabla\times\mathbf{H}=j\omega\epsilon(\mathbf{r})\epsilon_0\mathbf{E},
\end{equation}
where $\omega$ is the, yet to be determined, coupled frequency and $\epsilon\equiv\epsilon_r\epsilon_0$ is the dielectric constant determined by (\ref{eq:epsilon}).
Using the identity
\begin{equation}
\label{eq:VecIdent}
\nabla\cdot(\mathbf{A}\times \mathbf{B})=(\nabla\times \mathbf{A})\cdot\mathbf{B}-(\nabla\times \mathbf{B})\cdot\mathbf{A},
\end{equation}
(\ref{eq:expansionE}) and (\ref{eq:expansionH}), it can be shown that after integrating $\nabla \cdot(\mathbf{E}_i^*\times\mathbf{H})$ and $\nabla \cdot(\mathbf{E}\times\mathbf{H}_i^*)$ over some arbitrary volume $V$, two coupled equations in the coefficients $a$ and $b$ are obtained \cite{Elnaggar2015ECMT, Elnaggar2014coupled},
\begin{equation}
\label{eq:Coupled1}
\left(\Omega \mathcal{C}+j\mathcal{M}\right)b-\omega \mathcal{A}=0
\end{equation}
and
\begin{equation}
\label{eq:Coupled2}
-\omega\mathcal{C}b+\Omega\mathcal{D}a=0.
\end{equation}
Here $\Omega=\omega_0I$, $\mathcal{C}_{ik}=\int_{V}\mu_0\mathbf{H}_i^*\cdot\mathbf{H}_k dv$, $\mathcal{A}_{ik}=\int_{V}\epsilon\mathbf{E}_i^*\cdot\mathbf{E}_kdv$, $\mathcal{D}_{ik}=\int_{V}\epsilon_{i}\mathbf{E}_i^*\cdot\mathbf{E}_kdv$ and $\mathcal{M}_{ik}=\int_{\partial V}\mathbf{E}_i^*\times\mathbf{H}_k\cdot d\mathbf{S}$.

Starting from $\nabla\cdot\left(\mathbf{E}_i^*\times\mathbf{H}_k\right)$, using (\ref{eq:VecIdent}) and integrating over $V$ the boundary quantities $\mathcal{M}_{ik}$ terms can be expressed in terms of the bulk quantities $\mathcal{D}_{ik}$ and $\mathcal{C}_{ik}$ as
\begin{equation}
\label{eq:usefulRelation}
\mathcal{M}_{ik}=j\omega_0\left(\mathcal{C}_{ik}-\mathcal{D}_{ki}\right).
\end{equation}
Based on (\ref{eq:usefulRelation}), important relations can be found. Assuming that $V$ is taken to be all space, $\mathbf{E}_i$ and $\mathbf{H}_i$ vanish on the surface $\partial V$. Therefore,
\begin{equation}
\mathcal{C}_{ik}=\mathcal{D}_{ki}.
\end{equation}
When $i=k$, the above equation is consistent with the resonance condition, where the average stored magnetic and electric energies balance out. For $i\neq k$ and noting that $\mathcal{C}_{12}=\mathcal{C}_{21}$
\begin{equation}
\label{eq:reciprocal}
\mathcal{D}_{12}=\mathcal{D}_{21},
\end{equation}
an important property and will be used later to show that the interaction between DR1 and DR2 is reciprocal.
Eliminating $b$ from (\ref{eq:Coupled1}), (\ref{eq:Coupled2}) and using (\ref{eq:usefulRelation}), the frequency and fields of the coupled system can be determined from the solution of an eignevalue problem
\begin{equation}
\label{eq:eigenvalue}
\omega_0^2\mathcal{A}^{-1}\mathcal{D}\mathcal{C}^{-1}\mathcal{D}a=\omega^2a,
\end{equation}
identical to the eigenvalue problem derived in Ref. \onlinecite{Elnaggar2015ECMT}, which is not surprising since the procedure here parallels the one in Ref. \onlinecite{Elnaggar2015ECMT}. Noting that $\mathcal{D}_{ii}=\mathcal{C}_{ii}$, assuming that $\mathcal{A}_{11}\mathcal{A}_{22}\gg\mathcal{A}_{12}\mathcal{A}_{21}$, $\mathcal{C}_{11}\mathcal{C}{22}\gg\mathcal{C}_{12}\mathcal{C}_{21}$, $\mathcal{A}_{ii}\approx\mathcal{D}_{ii}$ and normalizing the modes such that $\mathcal{D}_{11}=\mathcal{D}_{22}=2$ (equivalently, energy of modes normalized to 1 Joule), (\ref{eq:eigenvalue}) is approximated to
\begin{equation}
\label{eq:evp1}
\omega_0^2
\begin{bmatrix}
1&-\frac{\left(\mathcal{A}_{12}-\mathcal{D}_{12}\right)}{2}\\
-\frac{\left(\mathcal{A}_{12}-\mathcal{D}_{21}\right)}{2} &1
\end{bmatrix}
\begin{bmatrix}
a_1\\a_2
\end{bmatrix}
=\omega^2
\begin{bmatrix}
a_1\\a_2.
\end{bmatrix},
\end{equation}
where (\ref{eq:reciprocal}) was used. The off-diagonal terms appearing in the above equation are equal. However they can be interpreted in two different, yet compatible, ways. The $(1,2) $ term can be written as
\begin{equation}
\label{eq:kappa1}
\kappa\equiv\frac{\mathcal{A}_{12}-\mathcal{D}_{12}}{2}=\frac{1}{2}\int_{DR2}\mathbf{P}_2\cdot\mathbf{E}_1dv,
\end{equation}
where $\mathbf{P}_2=\epsilon_0(\epsilon_{r2}-1)\mathbf{E}_2$ is the polarization vector of the DR2 mode.  Hence, $\kappa$ is the normalized maximum energy due to the interaction of the fields of DR1 with the polarization vector of DR2 and is identical to the expression previously obtained based on general physical arguments only \cite{wptc2, Elnaggar2015JAP}. The $(2,1)$ term, however, attains a different form
\begin{equation}
\label{eq:kappa2}
\kappa=\frac{\mathcal{A}_{12}-\mathcal{D}_{21}}{2}=\frac{1}{2}\int_{DR1}\left(\mathbf{P}_1\cdot\mathbf{E}_2-\mathbf{P}_2\cdot\mathbf{E}_1\right)dv.
\end{equation}
Relation (\ref{eq:reciprocal}) clearly shows that both expressions of $\kappa$ are mathematically identical. However (\ref{eq:kappa2}) can be interpreted as the \emph{net} of two normalized maximum energy terms: the stored energy in the polarization vector $\mathbf{P}_1$ due to its interaction with the aqueous field $\mathbf{E}_2$ minus the energy due to DR1 that would have existed in the displaced volume. As has been previously shown the solution of the EVP (\ref{eq:eigenvalue}) or (\ref{eq:evp1})  gives two coupled modes : the symmetric (bonding) with frequency $f_b$ and anti-symmetric (anti-bonding) with frequency $f_a>f_b$ \cite{Elnaggar2015ECMT, Elnaggar2014coupled, wptc2}.
\subsection{Coupling Coefficient Expressions}
In general DR1 is placed close to the surface. Since our interest here is in the interaction of the DRs $TE_{01\delta}$ modes, it is more convenient to first seek a closed form expression for $\kappa_0$, the situation depicted in Fig.\ref{fig:DRinDR} where DR1 is symmetrically placed inside DR2 (i.e, $z_0=0$). Noting that $l_1\ll l_2$, $\kappa$, when DR1 is displaced by a distance $d$ from DR2 centre, can be calculated as \cite{wptc2}
\begin{equation}
\label{eq:kappamisplaced}
\kappa=\kappa_0\cos(\beta_2z_0),
\end{equation}

The fields of DR1 and DR2 $TE_{01\delta}$ modes are determined using the Cohn model \cite{Pozar05}
\begin{equation}
\label{eq:CohnModel}
E_{\phi i}=M_i J_1(k_ir)\begin{cases}  \cos\beta_iz &\textnormal{\hspace{-9mm}}|z|\leq \frac{l_i}{2} \textnormal{ and } r\leq r_i\\
e^{\alpha_i\left(l_i/2-|z|\right)}\cos\frac{\beta_il_i}{2} & |z|>\frac{l_i}{2}\textnormal{ and } r\leq r_i\\
0 & \textnormal{otherwise}
\end{cases},
\end{equation}
where $r_i$, $l_i$ are the DR radius and length, respectively (Fig. \ref{fig:DRinDR}). The dielectric-air interface is assumed to be a PMC (Perfectly Magnetic Conductor) boundary, hence allowing the approximation of the radial wave number to $k_i=2.405/r_i$. The axial propagation and attenuation constants, $\beta_i$ and $\alpha_i$ are determined after solving the characteristic equation
\begin{equation}
\label{eq:drcharcteristiceqn}
\beta_i\tan\frac{\beta_i l_i}{2}-\alpha_i=0,
\end{equation}
given that $\beta_i=\sqrt{\epsilon_{ri}k_0^2-k_i^2}$ and $\alpha_i=\sqrt{k_i^2-k_0^2}$, where $k_0\equiv\omega/c$ is the free space wave number.
Although $\kappa$ can be calculated using either (\ref{eq:kappa1}) or (\ref{eq:kappa2}), the Cohn model does not consider the fields at $r>r_i$ to be relevant (ignores the diffraction of the fields, due to the PMC assumption). Hence, we exploit the reciprocal nature of $\kappa$ that was proved in Subsection II-A to deduce that (\ref{eq:kappa2}) is more appropriate to use to calculate $\kappa$ since it considers the interaction to be over the DR1 volume only. Accordingly,
\begin{equation}
\label{eq:kappa_expression}
\kappa_0=\frac{\left(\epsilon_{r1}-\epsilon_{r2}\right)\epsilon_0}{2}\int_{DR1}\mathbf{E}_1\cdot\mathbf{E}_2dv.
\end{equation}
Using (\ref{eq:CohnModel})
\begin{equation}
\kappa_0\approx \pi\epsilon_0(\epsilon_{r1}-\epsilon_{r2})M_1M_2I_1 \cdot I_2,
\end{equation}
where $I_1= \int_0^{r_1}rJ_1(k_1r)J_1(k_2r)dr$ and $I_2= \int_{-l_1/2}^{l_1/2}\cos\beta_1z\cos\beta_2zdz$.
Since $r_1\ll r_2$, $J_1(k_2r)\approx k_2r/2$ for $r\leq r_1$. Also
\begin{equation}
M_i=\frac{2}{\sqrt{\epsilon_{ri}\epsilon_0l_i^\textnormal{eff}}r_i},
\end{equation}
where $l_i^\textnormal{eff}=l_i+\sin\beta_il_i/\beta_i$ and $M_i$ is selected such that $\mathcal{D}_{ii}=2$. Therefore (\ref{eq:kappa_expression}) simplifies to
\begin{equation}
\label{eq:kappa}
\kappa_0\approx 1.6\sqrt{\frac{\epsilon_{r1}}{\epsilon_{r2}}}\left( \frac{r_1}{r_2}\right)^2\frac{l_1}{\sqrt{l_1^\textnormal{eff}l_2^\textnormal{eff}}}\left(S_++S_-\right),
\end{equation}
where $S_+=\textnormal{sinc}\left([\beta_1+\beta_2]l_1/2\right)$ and $S_-=\textnormal{sinc}\left([\beta_1-\beta_2]l_1/2\right)$. 

Additionally, $\kappa_0$ can be calculated from the coupled frequencies as
\begin{equation}
\label{eq:kappa3}
\kappa_0=\frac{f_a^2-f_b^2}{f_a^2+f_b^2}.
\end{equation}
Unlike (\ref{eq:kappa1}), (\ref{eq:kappa2}) and (\ref{eq:kappa}), (\ref{eq:kappa3}) is a \emph{phenomenological} relation that relates $\kappa_0$ to the observed coupled frequencies $f_b$ and $f_a$. 

\begin{figure}
\centering
\includegraphics[width=3.5in]{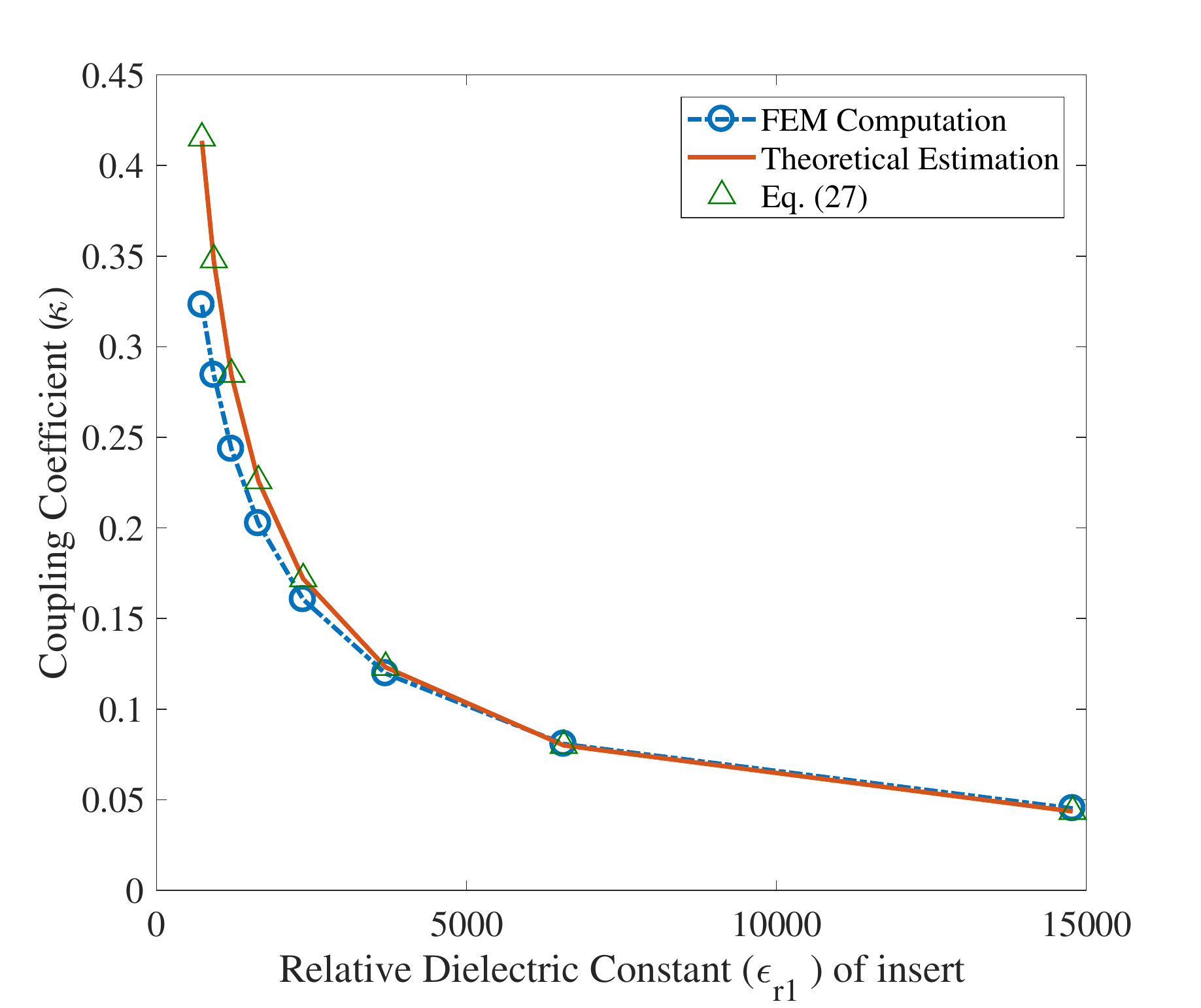}
\caption{The coupling coefficient $\kappa$ as a function of DR1 $\epsilon_r$.}
\label{fig:kappavsepsr}
\end{figure}

To verify the accuracy of (\ref{eq:kappa}), different numerical experiments are performed. The parameters of the containing DR (DR2) are fixed ($r_2=1.8 \textnormal{ m},~l_1=3\textnormal{ m}$ and $\epsilon_{r2}=~81$), while $\epsilon_{r1}$ and the dimensions of DR1 are allowed to change such that the resonant frequency is always fixed at 8.47 MHz. Different configurations are simulated using HFSS\textsuperscript\textregistered~Eigenmode solver, where $\kappa_0$ is calculated from the frequencies of the coupled modes (bonding and anti-bonding) as given by (\ref{eq:kappa3}). Additionally given the different parameters, $\beta_{1,2}$ and $l_{1,2}^\textnormal{eff}$ can be calculated from (\ref{eq:drcharcteristiceqn} and their values are plugged in \ref{eq:kappa} to calculate $\kappa_0$ according to (\ref{eq:kappa}). Figure (\ref{fig:kappavsepsr}) presents the results. It can be noted that for very large $\epsilon_{r1}$ values, the agreement is excellent. When $\epsilon_{r1}$ decreases, (\ref{eq:kappa}) over-estimates the value of $\kappa_0$. This is attributed to the increase in DR1 diameter, which renders the $J_1(k_2r)=k_2r/2$ approximation less accurate. Additionally, the PMC assumption of the Cohn model becomes less valid as $\epsilon_{r1}$ decreases. However, this has a secondary effect, since $\epsilon_{r1}$ value is still considerably large.

The dependency of $\kappa_0$ on the different parameters deserves further discussion. If $r_1$ and $l_1$ are scaled by a factor $s$ (i.e, $r_1\rightarrow sr_1$ and $l_1\rightarrow sl_1$), to keep the resonant frequency constant, the corresponding $\epsilon_{r1}$ should be changed to $\epsilon_{r1}/s^2$. The effect of scaling is reflected in the values of $\beta_1$ which changes to $\beta_1/s$ (or equivalently reducing the guided wavelength by $s$). Additionally, $l_1^\textnormal{eff}\rightarrow sl_1^\textnormal{eff}$. The overall effect of scaling DR1 results in a net inverse dependency on $\epsilon_{r1}$, i.e,
\begin{equation}
\label{eq:kappafitting}
\kappa_0=\kappa_{0n}\left(\frac{\epsilon_{r1n}}{\epsilon_{r}}\right)^{3/4},
\end{equation}
where the subscript $n$ identifies some reference configuration, taken here to be $\epsilon_{r1n}=2365$. Hence, the smaller the $\epsilon_{r1}$ value, the larger the DR1 diameter, the higher $\kappa_0$ is. Fig. \ref{fig:kappavsepsr} shows that (\ref{eq:kappafitting}) can be used instead of (\ref{eq:kappa}) to calculate $\kappa_0$ if $\kappa_{0n}$ is known. In the subsequent sections, two typical sets of DR1 parameters will be used. The first will be denoted by \emph{1X}  which has the following parameters $\epsilon_{r1}=2365$, $r_1=41.5$ cm, $l_1=29.6 $ cm, $f_0=8.47$ MHz, $\beta_1\approx 5.5\textnormal{ m}^{-1}$ and  $\alpha_1\approx 5.8  \textnormal{ m}^{-1}$. The second DR1 is denoted by \emph{2X} to emphasize that its linear dimensions are double those of the \emph{1X} DR (or equivalently $\epsilon_{r1}=591.25$). In all cases the aqueous solution has the following parameters: $\epsilon_{r2}=81$, $r_2=1.8$ m, $l_2=3$ m, $f_0=8.47$ MHz, $\beta_2\approx 0.72\textnormal{ m}^{-1}$  and $\alpha_2\approx  1.32\textnormal{ m}^{-1}$.

Eq. (\ref{eq:kappafitting}) implies that the requirement $\epsilon_{r1}\gg\epsilon_{r2}$ can be relaxed if one uses a larger dielectric insert. In fact, using a \emph{2X} DR allows $\epsilon_{r1}$ to be reduced by a factor of four. Hence it is possible to trade-off DR dimension and its dielectric constant. Additionally it should be noted that, everything else fixed, a lower $\epsilon_{r1}$ value implies a higher $\kappa$, which in turn translates into a more isolation of the non-bonding mode and higher transfer efficiency \cite{Elnaggar2017JAP}. Unfortunately, the reduction of $\epsilon_{r1}$ comes at the cost of an increase in the DR intrinsic losses, as will be shown in Subsection III-B.

\section{Interaction of two DRs with the aqueous medium}
\begin{figure}
\centering
\includegraphics[width=2.5in]{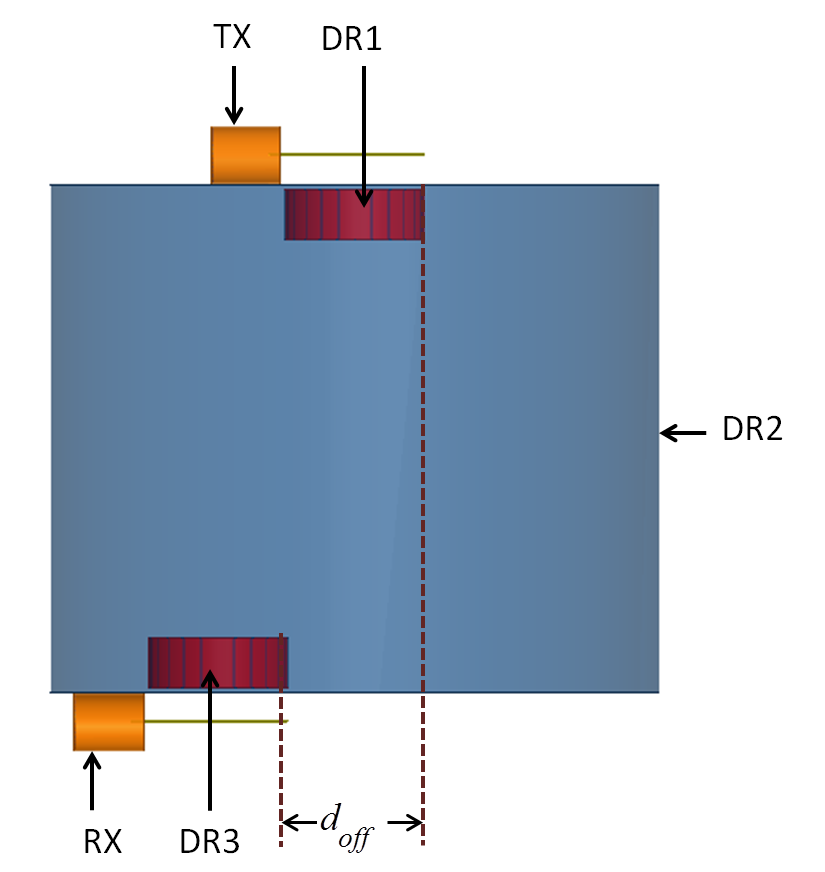}
\caption{Two ultra high $\epsilon_r$ DR resonators (DR1 and DR2) inserted in a contained volume of a high $\epsilon_r$ aqueous solution DR2. The three resonators modes have the same resonant frequency. At the transmitting and receiving ends, power is coupled into and out via two loops.}
\label{fig:SystemConfig}
\end{figure}
In principle, it is possible to extend the CMT analysis of the previous section to the case where two DRs (DR1 and DR3) are present in addition to DR2. In fact, such generalization of the CMT approach to include an arbitrary number of modes was previously reported \cite{Elnaggar2015ECMT} and extended to situations where resonators are in the vicinity of PEC or PMC boundaries \cite{Elnaggar2015Image}. In the following treatment we will take advantage of the already developed $2\times 2$ eigenvalue problem (\ref{eq:eigenvalue}) and directly extend it to to the $3\times 3$ case by populating the off-diagonals with the necessary elements. It is assumed that DR1 and DR3 are far enough such that their interaction can be neglected. This is not quite true, particularly when the \emph{2X} DRs are deployed. The \emph{direct} interaction between DR1 and DR3 will improve the efficiency as it opens a direct channel, in parallel to the \emph{indirect} interaction via the aquaeous modes. Furthermore, the general asymmetric case (i.e, $\kappa_{12}\neq\kappa_{32}$), arising from the misalignment of DR1 and DR3, is considered. Such situation may naturally arise in the present configuration due to inevitable turbulences and drift currents that may exist in the medium. Accordingly the $3\times 3$ EVP is simplified to
\begin{equation}
\label{eq:eigenvalue3Res}
\begin{bmatrix}
\omega_0^2 &-\omega_0^2\kappa_{12} &0\\
-\omega_0^2\kappa_{12} & -\omega_0^2 &-\omega_0^2\kappa_{32}\\
0 & -\omega_0^2\kappa_{32} &\omega_0^2
\end{bmatrix}
\begin{bmatrix}
a_1\\a_2\\a_3
\end{bmatrix}
=\omega^2
\begin{bmatrix}
a_1\\a_2\\a_3
\end{bmatrix},
\end{equation}
where $\kappa_{12}$ and $\kappa_{32}$ are the coupling coefficients between DR1 and DR2, and DR3 and DR3, respectively. The coupled modes are the eigensolutions of (\ref{eq:eigenvalue3Res}), which take the form
\begin{equation}
\label{eq:bonding}
\omega_b=\omega_0\sqrt{1-\left(\kappa_{12}^2+\kappa_{32}^2\right)^{1/2}},
\end{equation}
\begin{equation}
\label{eq:bondingV}
\mathbf{V}_b=\frac{1}{\sqrt{2}}\left[\frac{\zeta}{\sqrt{1+\zeta^2}}, 1, \frac{1}{\sqrt{1+\zeta^2}}\right]^t
\end{equation}
for the bonding mode,
\begin{equation}
\label{eq:nonbonding}
\omega_n=\omega_0,~~\mathbf{V}_n=\left[-\frac{1}{\sqrt{1+\zeta^2}},0, \frac{\zeta}{\sqrt{1+\zeta^2}}\right]^t
\end{equation}
for the non-bonding mode, and
\begin{equation}
\label{eq:antibonding}
\omega_a=\omega_0\sqrt{1+\left(\kappa_{12}^2+\kappa_{32}^2\right)^{1/2}},
\end{equation}
\begin{equation}
\label{eq:antibondingV}
\mathbf{V}_a=\frac{1}{\sqrt{2}}\left[\frac{\zeta}{\sqrt{1+\zeta^2}}, -1, \frac{1}{\sqrt{1+\zeta^2}}\right]^t
\end{equation}
\begin{figure}
\centering
\includegraphics[width=3.0in, angle=5]{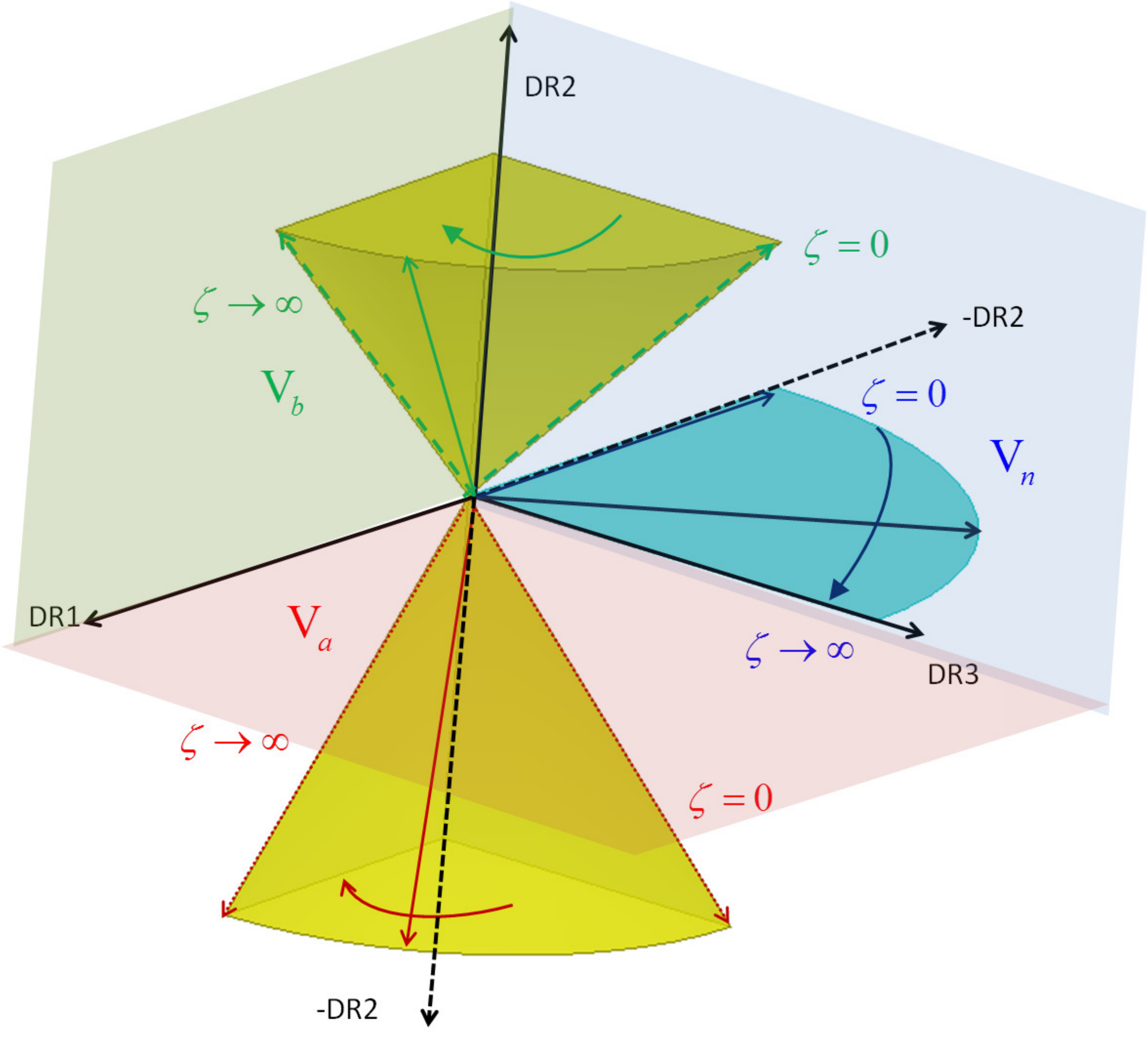}
\caption{(a) 3D Space of Eigenvectors. Space of Bonding, Non-bonding and  Anti-bonding modes for $0\leq\zeta <\infty$.}
\label{fig:vectors3d}
\end{figure}
for the anti-bonding mode. Here, $\zeta\equiv\kappa_{12}/\kappa_{32}$.  For convenience,  all modes are normalized such that the eigenvectors in (\ref{eq:bondingV}), (\ref{eq:nonbonding}) and (\ref{eq:antibondingV}) are unit vectors. The parameter $\zeta$ changes from zero (no coupling between DR1 and DR2), to one (equal coupling, $\kappa_{12}=\kappa_{32}$), to infinity (no coupling between DR3 and DR1). The eigen-frequencies and the corresponding eigenvectors change as functions of $\zeta$. Fig. \ref{fig:vectors3d} depicts the different locations of the eigenvectors. One interesting property is that, under the assumptions of negligible Coupling Induced Frequency Shifts (CIFS)\cite{Popovic06} that appear as on diagonal terms and $\kappa_{13}=0$, the frequency of the non-bonding mode is fixed at $\omega_0$ and its eigenvector is always in the DR1 and DR3 plane with no DR2 component. As expected, under symmetric alignment (i.e, $\kappa_{12}=\kappa_{32}$) the DR1 and DR3 modes contribute equally to the non-bonding mode and they are $180^\circ$ out of phase. Fig. \ref{fig:vectors3d} also shows that the eigenvector of the bonding (anti-bonding) mode inscribes a quarter of a cone around the positive (negative) DR2 mode as $\zeta$ changes from 0 to $\infty$ as $\zeta$ changes from zero to infinity. The extreme case $\zeta=0$ ($\zeta\rightarrow\infty$) describes the limiting situation at which the non-bonding mode represents the mode of DR1 (DR3) and the bonding and anti-bonding modes are the symmetric and anti-symmetric modes, respectively due to the coupling of DR2 and DR3 (DR1). 

\begin{figure*}
\centering
\includegraphics[width=6.0in]{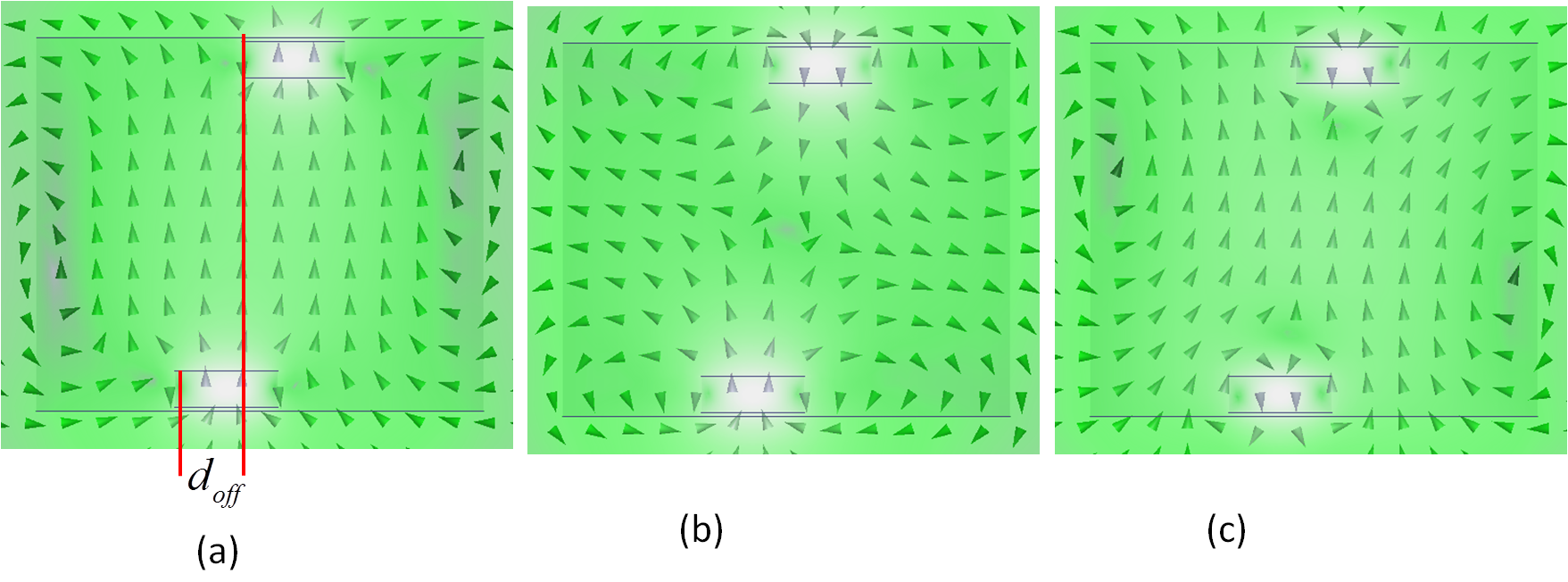}
\caption{Magnetic field of (a) Bonding ($f_b=7.95~\textnormal{MHz}$)  (b) Non Bonding ($f_n=8.3 \textnormal{ MHz}$) and  (c) Anti-bonding ($f_a=9.1 \textnormal{ MHz}$) modes  when DR1 and DR3 are displaced by 27 cm from DR2 axis.}
\label{fig:modes}
\end{figure*}

To show that the three coupled modes do exist in the general case when DR1 and DR3 are misaligned, the structure in Fig. \ref{fig:SystemConfig} is simulated where both resonators are displaced by 27 cm away from DR axis. The magnetic field profile is depicted in Fig. \ref{fig:modes}. The frequency of the non-bonding mode has been shifted down from 8.47 MHz to 8.3 MHz, which can be attributed to the effect of coupling with other modes, not taken into account in the previous analysis and that may overlap with the DRs $TE_{01\delta}$ modes due to the axial eccentricity. Inspection of the fields reveals that the relative phases of the different components (fields of DR modes) do agree with the eigenvector expressions (\ref{eq:bondingV}), (\ref{eq:nonbonding}), (\ref{eq:antibondingV}) and Fig. \ref{fig:vectors3d}. For the anti-bonding mode (Fig. \ref{fig:modes}(b)) for instance, the DR2 mode is $180^\circ$ out of phase with both DR1 and DR3 modes, as (\ref{eq:antibondingV}) predicts (or equivalently, $\mathbf{V}_a$ is restricted to the DR1-DR3 plane).

\subsection{Modal Expansion and the excitation of the non-bonding mode}
The complete set of modes of a given system fully describes its dynamical behaviour. For sinusoidal excitation with frequency $\omega$, the response is the weighted summation of all eigenmodes i.e,
\begin{equation}
\label{eq:modalexpansion}
\tilde{Y}(\omega)=\sum_{k=1}^\infty \frac{A_k}{\omega-\omega_k-i\sigma_k}+c.c,
\end{equation}
where $\omega_k-i\sigma_k$ is complex frequency of the $\textnormal{k}^\textnormal{th}$ mode, and $A_k$, the expansion coefficient, depends mainly on the coupling of the input excitation to the given mode. To excite the mode, it is thus desirable that there is an overlap between its profile and the source to assure that $A_k$ is sufficiently large. For high $Q$ systems, as in our case here, $\sigma_k$ is small. When $\omega$ coincides with one of the system frequencies $\omega_k$, its contribution to (\ref{eq:modalexpansion}) becomes dominant; this is particularly true when the frequency of other modes are sufficiently far from the given mode. The considerably strong coupling between the modes in the DRs-Aqueous situation allows the modes to be well separated to the extent that the non-bonding mode prevails as the excitation frequency gets close to the resonant frequency $\omega_0$.

\subsection{Reduction of Intrinsic DR $Q_0$ due to the aqueous medium}
\begin{figure}
\centering
\includegraphics[width=3.5in]{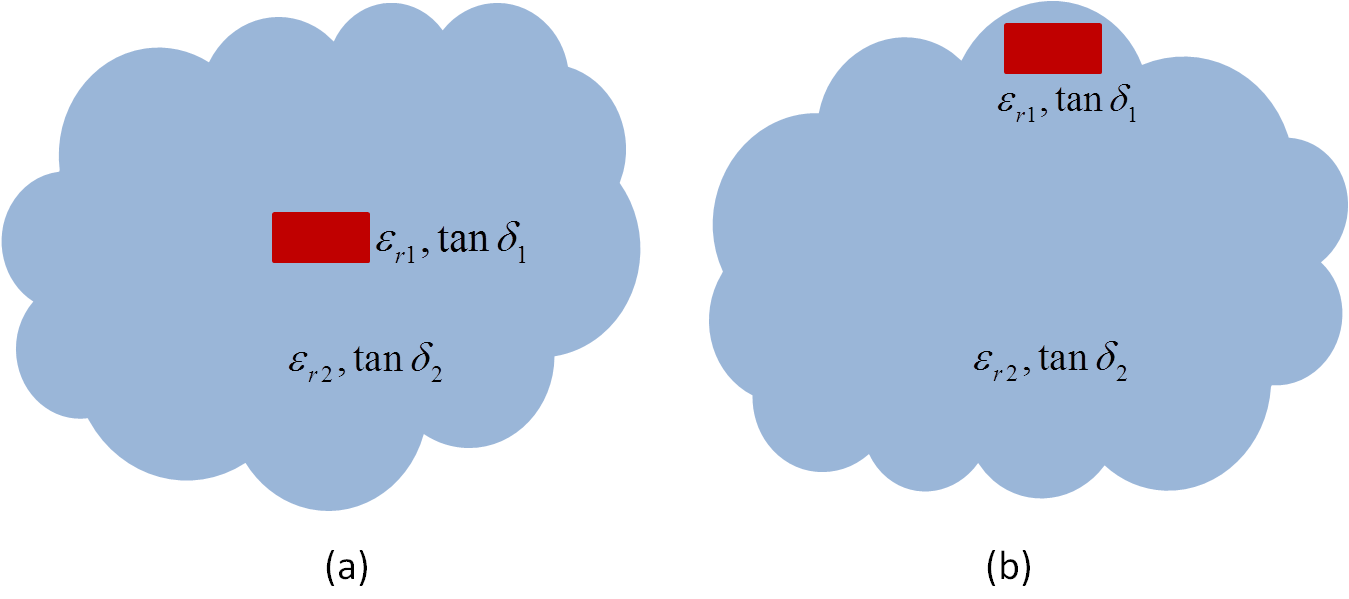}
\caption{(a) DR1 inside the aqua solution DR2. (b) DR1 close to the surface of DR2.}
\label{fig:appendix}
\end{figure}
The intrinsic losses of DR1 or DR3 are the losses of the mode in the absence of the load.  Neglecting radiation for such high $\epsilon_r$ resonators, the losses are mainly due to the dielectrics loss tangents. The Quality factor is defined by $Q_0\equiv 2\omega_0W_E/P_l$, where $W_E$ is the average stored electric energy (equals to the average stored magnetic energy at resonance) and $P_l$ is the average power loss. The total electric energy $W_E$ is the sum of the energy inside and outside DR1 (i.e, $W_E=W_E^\textnormal{in}+W_E^\textnormal{out}$).

The power loss is given by
\begin{equation}
P_{l}=P_{l}^\textnormal{in}+P_{l}^\textnormal{out},
\end{equation}
where $P_{l}^\textnormal{in}$ is the power loss inside DR1 (DR3) material and $P_{l}^\textnormal{out}$ denotes the losses due to the DR1 (DR3) fields fringing into DR2. Therefore,
\begin{equation}
P_{l}^\textnormal{in}=2\omega_0\tan\delta_1W_E^\textnormal{in}
\end{equation}
and
\begin{equation}
P_{l}^\textnormal{out}=2\omega_0\tan\delta_2W_E^\textnormal{out}.
\end{equation}
For the configuration in Fig. \ref{fig:appendix}(b), the losses are approximately half that in Fig. \ref{fig:appendix}(a). If $x$ is defined as the ratio of $W_E^\textnormal{out}$ and $W_E^\textnormal{in}$, it is readily found that
\begin{equation}
\label{eq:Q0}
Q_0=\frac{1+x}{\tan\delta_1+\gamma x\tan\delta_2},
\end{equation}
where $\gamma=1$ and 0.5 for the configurations in Fig. \ref{fig:appendix} (a) and Fig. \ref{fig:appendix}(b), respectively.
Note that $W_E^\textnormal{in}$ and $W_E^\textnormal{out}$ are proportional to the integral of the square of $\mathbf{E}$ over the corresponding volumes,
\begin{equation}
W_E^\textnormal{in}=\frac{1}{2}\int_\textnormal{in}\epsilon_{r1}|\mathbf{E}|^2dV
\end{equation}
and
\begin{equation}
\label{eq:WEout}
W_E^\textnormal{out}=\frac{1}{2}\int_\textnormal{out}\epsilon_{r2}|\mathbf{E}|^2dV,
\end{equation}
For the \emph{1X} DR1, $x\approx 0.25$, while it increases to approximately 0.43 for the \emph{2X} DR. It is always desirable that $\tan\delta_1$ is as small as possible to avoid any unnecessary losses. In this case $\gamma x\tan\delta_2\gg \tan\delta_1$ hence (\ref{eq:Q0}) reduces to
\begin{equation}
\label{eq:Q0simplified}
Q_0\approx\frac{1+x}{\gamma x\tan\delta_2},
\end{equation}
which is independent of $\tan\delta_1$. For a given medium, $Q_0$ decreases as more fields fringe outside DR1, which is the case when $\epsilon_{r1}$ is reduced. Additionally, (\ref{eq:Q0simplified}) has an interesting implication; it was previously shown that the efficiency $\eta$ attains its maximum when the $Q$ of the load is \cite{Elnaggar2017JAP}
\begin{equation}
\label{eq:Q_max}
Q_w^\textnormal{max}\approx\frac{1}{\kappa}\sqrt{\frac{Q_0}{Q_2}}.
\end{equation}
Substituting (\ref{eq:Q0simplified}) in (\ref{eq:Q_max}) shows that $Q_w^\textnormal{max}$ is independent of the medium loss tangent. The maximum efficiency $\eta^\textnormal{max}$, of course, will decrease as the medium losses increase. The maximum efficiency $\eta^\textnormal{max}$ is given by \cite{Elnaggar2017JAP}
\begin{equation}
\label{eq:etamax}
\eta^\textnormal{max}\approx 100\left(1-\frac{2}{\sqrt{\kappa Q_0}}\right),
\end{equation}
a function of $x$ and the medium loss tangent $\tan\delta_2$.

\begin{figure}
\centering
\includegraphics[width=3.0in]{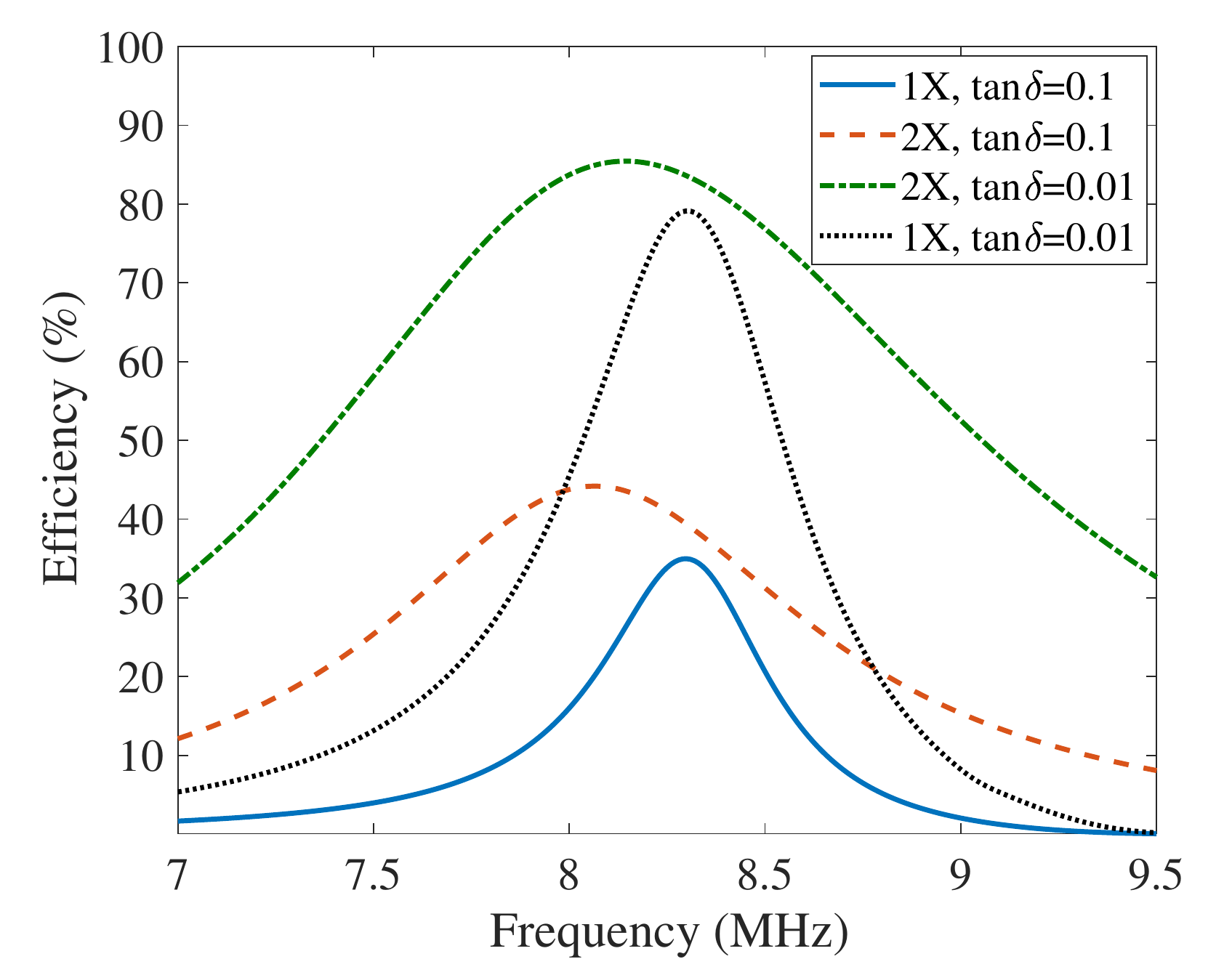}
\caption{Efficiency comparison when using \emph{1X} and \emph{2X} sized DRs for two different DR2 loss tangent values.}
\label{fig:1xvs2x}
\end{figure}

To demonstrate the effect of up-sizing DR1 and DR3 on $\eta$, the maximum efficiency is calculated for both the \emph{1X} and \emph{2X} structures when placed inside the same aqueous solution. The efficiency versus frequency is plotted in Fig. \ref{fig:1xvs2x}. According to (\ref{eq:kappafitting}) when DR1 linear dimensions are doubled, $\kappa\approx 2.8\kappa_n$. As shown in the Fig., the increase in $\kappa$ results in a larger separation of the coupled modes frequencies, which in turn extends the bandwidth. Additionally, the net effect of $\kappa Q_0$ is to increase $\eta$. It is worth noting that the frequency of the non-bonding mode has been shifted down due to the effect of the CIFS terms, which become significant for large values of $\kappa$ \cite{Popovic06,Elnaggar2015ECMT}. Unlike the \emph{1X} configuration, DR1 and DR3 in the \emph{2X} case directly couple to one another, as well as, coupling to the medium; this is similar to the situation previously observed in electron spin resonance probes \cite{AMR2017}.

Equation (\ref{eq:etamax}) shows that $\eta^\textnormal{max}$ depends on the product $\kappa Q_0$. Since the configurations studied here assume that DR1 and DR3 are close to the surface, $\gamma=0.5$. Therefore from (\ref{eq:Q0simplified}) $Q_0=100$ and 66.5 for the \emph{1X} and \emph{2X} DRs, respectively when $\tan\delta_2=0.1$. Noting that $\kappa$ at the surface is approximately 0.1 \cite{wptc2} and that $\kappa_{2X}=2.8\kappa_{1X}$, the maximum efficiency $\eta^\textnormal{max}$ values are 36.8\% (\emph{1X} and 53.6\% (\emph{2X}), which generally agree with the full-wave simulations reported in Fig. \ref{fig:1xvs2x}.

\begin{figure}
\centering
\includegraphics[width=3.0in]{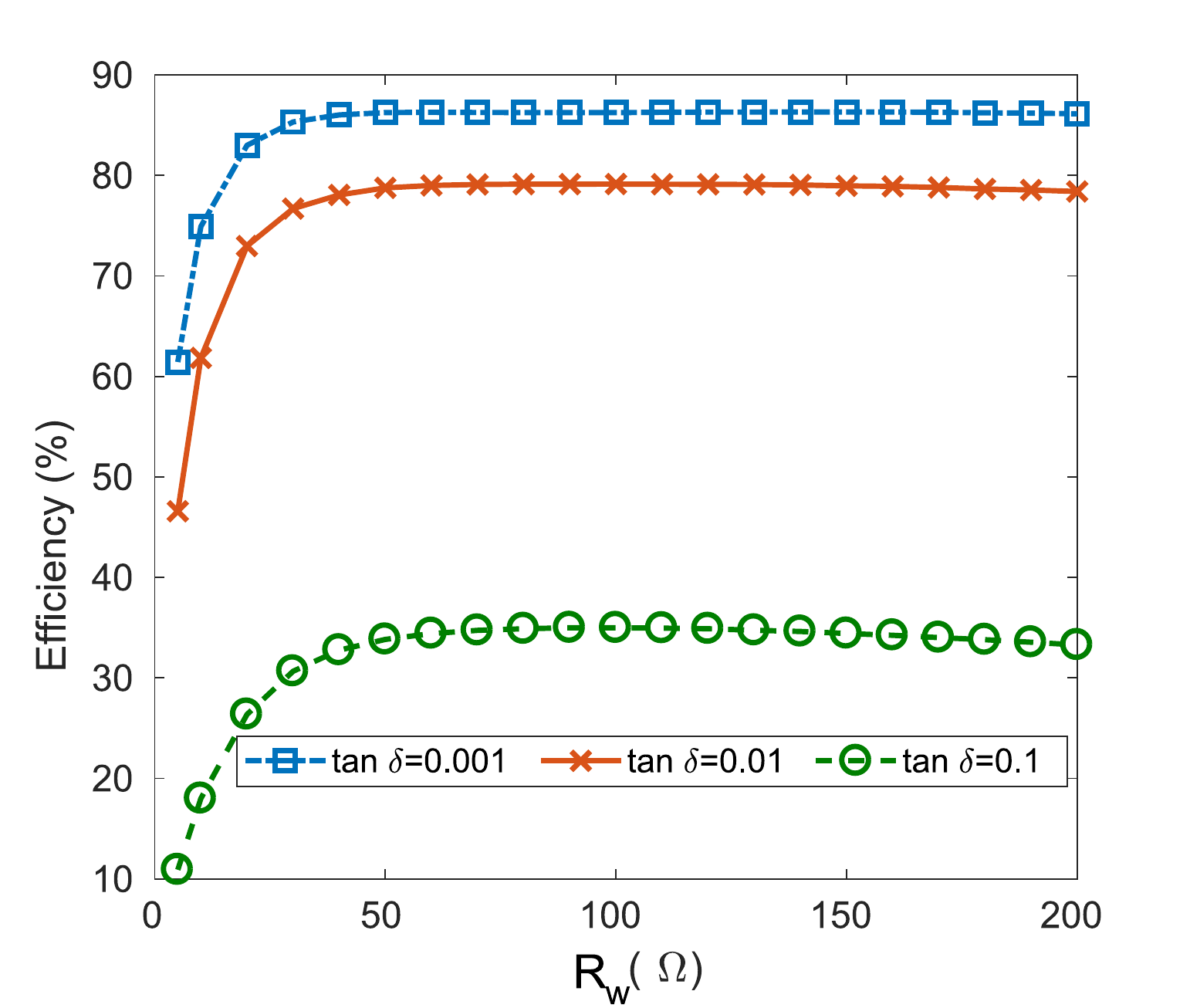}
\caption{Efficiency ($\eta$) as a function of the load impedance for different medium loss tangent $\tan\delta_2$.}
\label{fig:etaVsZ}
\end{figure}

To show that $Q_w^\textnormal{max}$ weakly depends on $\tan\delta_2$, the load is modelled as a lumped (discrete) port with some real impedance $R_w$. Increasing $R_w$ is equivalent to decreasing the load $Q_w$. The whole structure in Fig. \ref{fig:SystemConfig} is simulated using HFSS\textsuperscript\textregistered~for different $R_w$ and $\tan\delta_2$. Fig. \ref{fig:etaVsZ} presents the simulation results for three different $\tan\delta_2$. As can be seen from the Fig., the impedance at which $\eta$ attains its maximum is in the vicinity of $50~{\Omega}$ for all three $\tan\delta_2$ values, changing over two decades.

\begin{figure}
\centering
\includegraphics[width=3.0in]{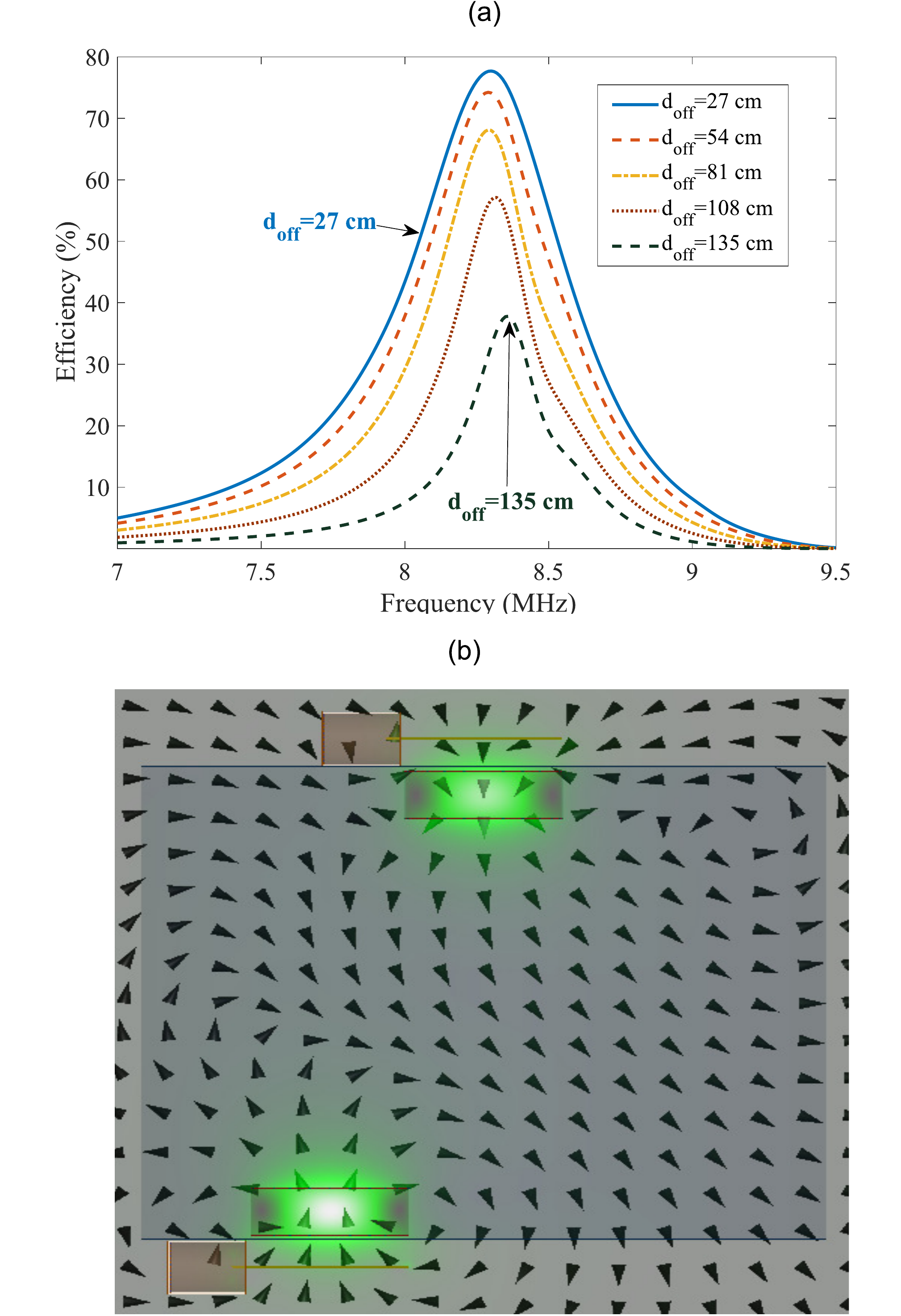}
\caption{(a) Efficiency vs. frequency for different values of $d_\textnormal{off}$. (b) The magnetic field profile at the frequency corresponding to the maximum efficiency for $d_\textnormal{off}=$ 81 cm.}
\label{fig:offset}
\end{figure}

The presence of the non-bonding mode over a wide range of $\zeta$ values can be exploited to enable efficient WPT, the offset between the Tx and Rx in Fig. \ref{fig:SystemConfig} can be substantial due to the presence of turbulences, drift currents, or possibly requiring Rx to freely move inside the medium (for instance Rx is attached to a swimmer inside a swimming pool). To demonstrate that efficient power transmission is possible over a wide range of $d_\textnormal{off}$ (Fig. \ref{fig:SystemConfig}), $d_\textnormal{off}$ is allowed to increase, resulting in a reduction of $\kappa_{32}$ . The whole system was simulated for the \emph{1X} resonators. As Fig.\ref{fig:offset}(a) presents, the efficiency does decrease as the offset increases. This is due to the decrease of coupling between DR3, the receiver, and DR2. For offsets even greater than 1 m the efficiency is still well above 50\%. Moreover, the maximum efficiency occurs at the frequency of the non-bonding mode, which is independent of the offset level, emphasizing the fact that the energy transfer is mainly due to the presence of the non-bonding mode. Figure \ref{fig:offset}(b) confirms that the field profile is indeed that of the non-bonding mode. For applications where turbulences and drift currents may occur, or where the alignment between DR1 and DR3 cannot be always maintained, the presence of the non-bonding mode enables power transfer to still be possible even in less than ideal situations.

The requirement and the possibility of having high $\kappa_{12}$ and $\kappa_{32}$ values along with the absence of the DR2 mode from the non-bonding mode (or equivalently the eigenvector always being in the DR1-DR3 plane, independent of $\zeta$),  are the two main desirable features that enable EIT-like wireless power transmission. The implication of the second property is briefly discussed in subsection \ref{subsect:disturbance}.

\subsection{Tolerating Conducting and Dielectric Objects inside the Aquaeous Solution}
\label{subsect:disturbance}

\begin{figure}
\centering
\includegraphics[width=3.0in]{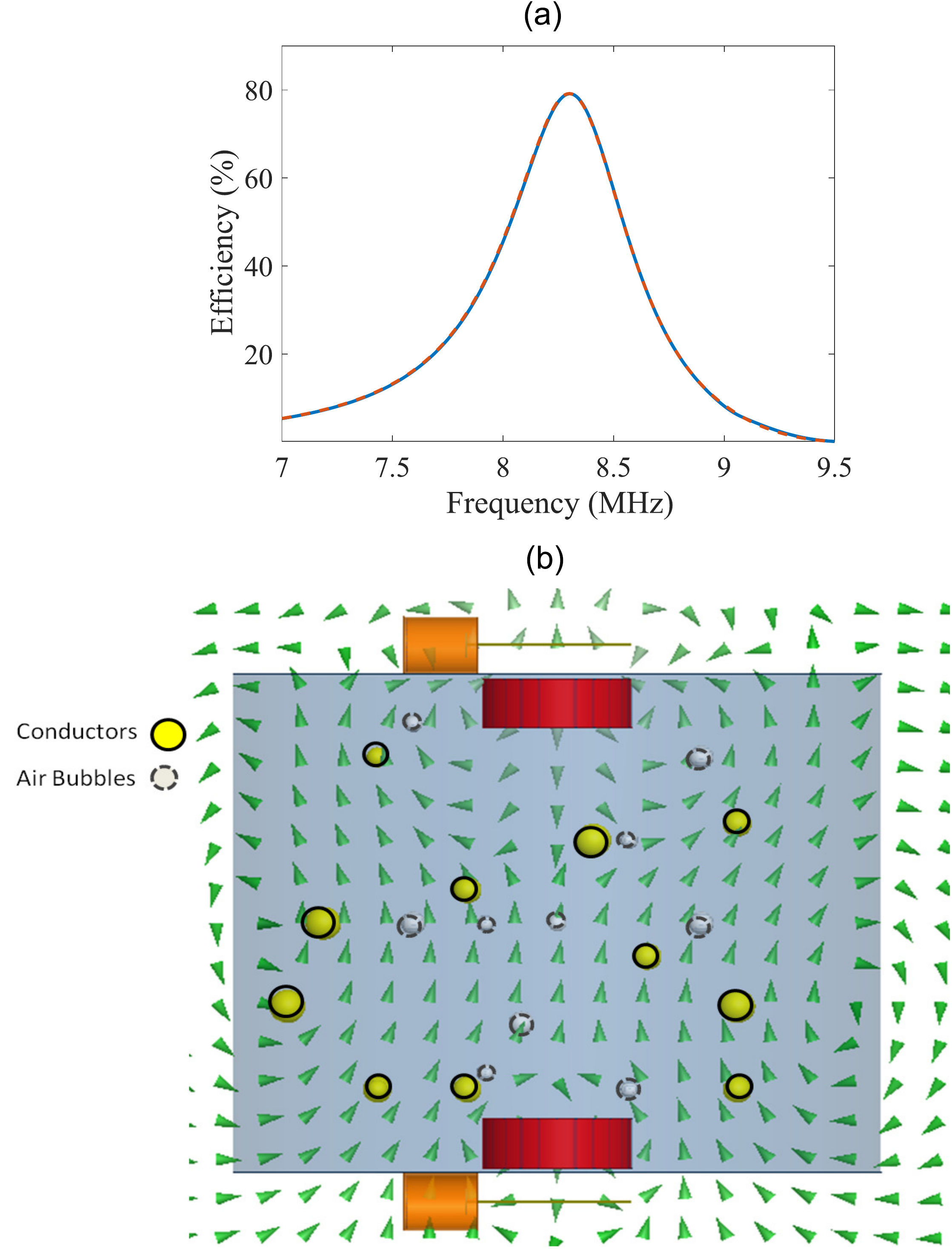}
\caption{Addition of randomly placed air bubbles and conducting objects. Change in efficiency is almost unobservable.}
\label{fig:bubbles}
\end{figure}

As applications may dictate, disturbances and objects may randomly appear in the container. For instance, air bubbles, rocks and other conducting elements such as a stirrer or mixer, etc, depending on the particular application, may interact with the proposed WPT system. The presence of the non-bonding mode and the high $\kappa$ values enable the medium disturbances to be of a secondary nature. To mimic a dynamic environment, particles that model air bubbles and conducting objects are randomly placed inside the container as Fig. \ref{fig:bubbles} shows. Interestingly enough, the change of efficiency is too small to observe, emphasizing that EIT-like WPT enabled high efficiency even in the presence of random extraneous objects.

\begin{figure}[!t]
\centering
\includegraphics[width=3.0in]{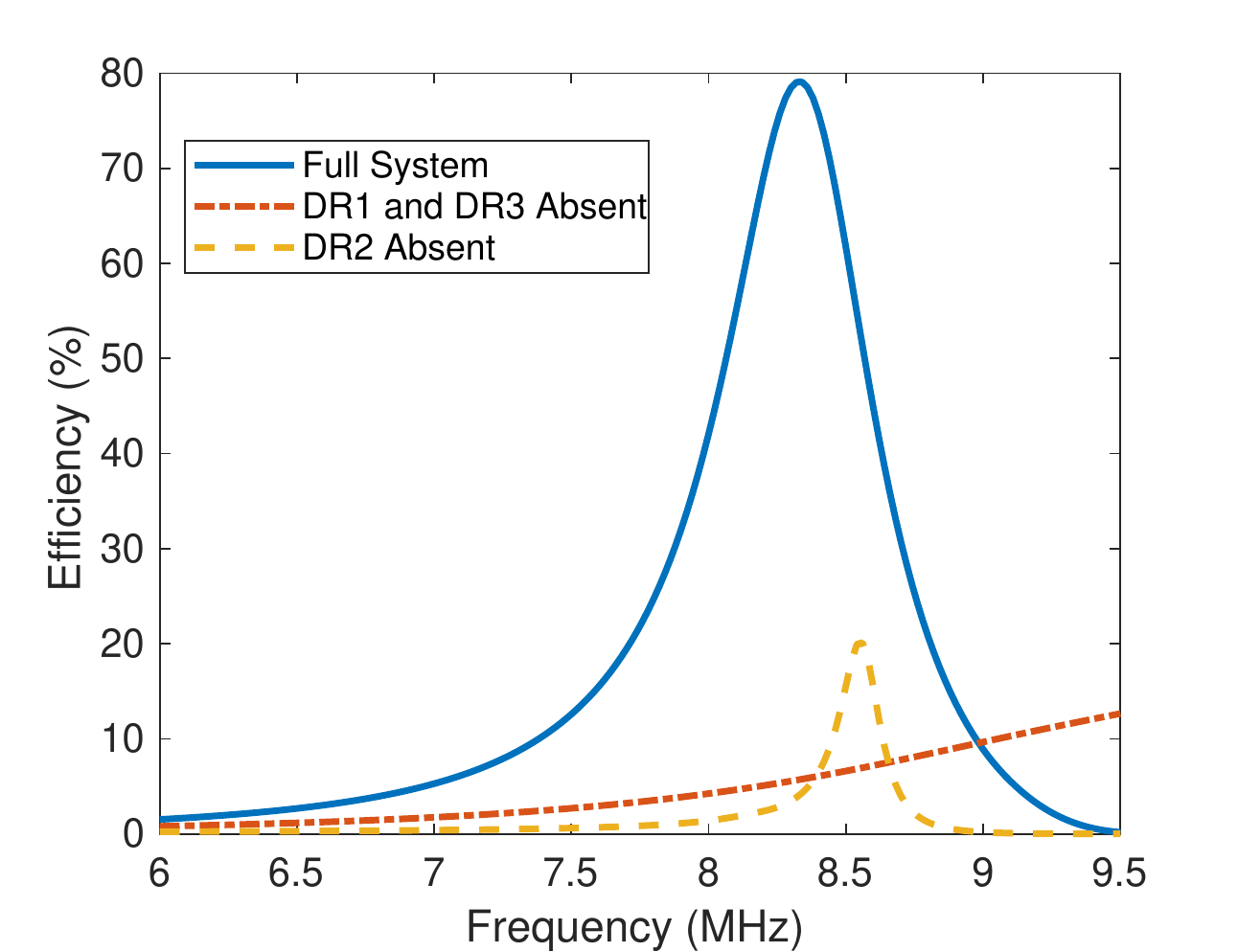}
\caption{Efficiency $\eta$ versus the excitation frequency for three different scenarios. DR2, when present, has a loss tangent $\tan\delta=0.01$.}
\label{fig:Efficiency0p01}
\end{figure}

 It is worth noting that the absence of the DR2 mode does not imply that the presence of DR2 is unnecessary. In fact, DR2 presence is vital for the whole transfer process to take place; it mediates the indirect interaction between DR1 and DR3. To show that the presence of all three elements, DR1, DR2 and DR3 is essential, the transfer efficiency is computed for three different configurations as shown in Fig. \ref{fig:Efficiency0p01}. When DR2 is absent, DR1 and DR3 are weakly coupled and the efficiency is around 20\%. If the aqueous medium is present and DR1 and DR3 are removed, power cannot be transferred except through DR2 modes, which is small since the excitation is applied to the container surface. When the three components coexist, efficiency is considerably increased as clearly revealed in Fig. \ref{fig:Efficiency0p01}.

\section{Conclusion}
In the current article, a systematic analysis of the interaction between a contained aqueous solution mode and dielectric resonators inserts is done. It is shown that the restriction on using high $\epsilon_r$ DRs can be relaxed. Lower $\epsilon_r$ resonators will strongly couple with the contained medium and hence improve the overall system efficiency. However, this comes with the expense of being potentially bulky. Additionally, the leakage electric field increases with the reduction of $\epsilon_r$ resulting in a reduction of the intrinsic $Q_0$. Fortunately, the lower the $\epsilon_r$ is the lower $\tan\delta_1$ will possibly be. The reciprocal nature of interaction emerges directly from the coupled mode equations and results in a symmetric eigenvalue operator. The general eigenvalue problem between two dielectric resonators and the aqueous medium were developed and solved. Pictorial presentation of the modes were proposed. The effect of misalignment on the transfer efficiency was examined and shown to be consistent with the behaviour of the modes. The possible excitation of the non-bonding mode with high fidelity, as well as robustness of the system against the presence of extraneous objects were also discussed.

\bibliography{WPTC2018}

\begin{thebibliography}{28}%
\makeatletter
\providecommand \@ifxundefined [1]{%
 \@ifx{#1\undefined}
}%
\providecommand \@ifnum [1]{%
 \ifnum #1\expandafter \@firstoftwo
 \else \expandafter \@secondoftwo
 \fi
}%
\providecommand \@ifx [1]{%
 \ifx #1\expandafter \@firstoftwo
 \else \expandafter \@secondoftwo
 \fi
}%
\providecommand \natexlab [1]{#1}%
\providecommand \enquote  [1]{``#1''}%
\providecommand \bibnamefont  [1]{#1}%
\providecommand \bibfnamefont [1]{#1}%
\providecommand \citenamefont [1]{#1}%
\providecommand \href@noop [0]{\@secondoftwo}%
\providecommand \href [0]{\begingroup \@sanitize@url \@href}%
\providecommand \@href[1]{\@@startlink{#1}\@@href}%
\providecommand \@@href[1]{\endgroup#1\@@endlink}%
\providecommand \@sanitize@url [0]{\catcode `\\12\catcode `\$12\catcode
  `\&12\catcode `\#12\catcode `\^12\catcode `\_12\catcode `\%12\relax}%
\providecommand \@@startlink[1]{}%
\providecommand \@@endlink[0]{}%
\providecommand \url  [0]{\begingroup\@sanitize@url \@url }%
\providecommand \@url [1]{\endgroup\@href {#1}{\urlprefix }}%
\providecommand \urlprefix  [0]{URL }%
\providecommand \Eprint [0]{\href }%
\providecommand \doibase [0]{http://dx.doi.org/}%
\providecommand \selectlanguage [0]{\@gobble}%
\providecommand \bibinfo  [0]{\@secondoftwo}%
\providecommand \bibfield  [0]{\@secondoftwo}%
\providecommand \translation [1]{[#1]}%
\providecommand \BibitemOpen [0]{}%
\providecommand \bibitemStop [0]{}%
\providecommand \bibitemNoStop [0]{.\EOS\space}%
\providecommand \EOS [0]{\spacefactor3000\relax}%
\providecommand \BibitemShut  [1]{\csname bibitem#1\endcsname}%
\let\auto@bib@innerbib\@empty
\bibitem [{\citenamefont {Hui}(2013)}]{hui2013planar}%
  \BibitemOpen
  \bibfield  {author} {\bibinfo {author} {\bibfnamefont {S.}~\bibnamefont
  {Hui}},\ }\href@noop {} {\bibfield  {journal} {\bibinfo  {journal}
  {Proceedings of the IEEE}\ }\textbf {\bibinfo {volume} {101}},\ \bibinfo
  {pages} {1290} (\bibinfo {year} {2013})}\BibitemShut {NoStop}%
\bibitem [{\citenamefont {Shin}\ \emph {et~al.}(2014)\citenamefont {Shin},
  \citenamefont {Shin}, \citenamefont {Kim}, \citenamefont {Ahn}, \citenamefont
  {Lee}, \citenamefont {Jung}, \citenamefont {Jeon},\ and\ \citenamefont
  {Cho}}]{shin2014design}%
  \BibitemOpen
  \bibfield  {author} {\bibinfo {author} {\bibfnamefont {J.}~\bibnamefont
  {Shin}}, \bibinfo {author} {\bibfnamefont {S.}~\bibnamefont {Shin}}, \bibinfo
  {author} {\bibfnamefont {Y.}~\bibnamefont {Kim}}, \bibinfo {author}
  {\bibfnamefont {S.}~\bibnamefont {Ahn}}, \bibinfo {author} {\bibfnamefont
  {S.}~\bibnamefont {Lee}}, \bibinfo {author} {\bibfnamefont {G.}~\bibnamefont
  {Jung}}, \bibinfo {author} {\bibfnamefont {S.-J.}\ \bibnamefont {Jeon}}, \
  and\ \bibinfo {author} {\bibfnamefont {D.-H.}\ \bibnamefont {Cho}},\
  }\href@noop {} {\bibfield  {journal} {\bibinfo  {journal} {IEEE Transactions
  on Industrial electronics}\ }\textbf {\bibinfo {volume} {61}},\ \bibinfo
  {pages} {1179} (\bibinfo {year} {2014})}\BibitemShut {NoStop}%
\bibitem [{\citenamefont {Xue}, \citenamefont {Cheng},\ and\ \citenamefont
  {Je}(2013)}]{xue2013high}%
  \BibitemOpen
  \bibfield  {author} {\bibinfo {author} {\bibfnamefont {R.-F.}\ \bibnamefont
  {Xue}}, \bibinfo {author} {\bibfnamefont {K.-W.}\ \bibnamefont {Cheng}}, \
  and\ \bibinfo {author} {\bibfnamefont {M.}~\bibnamefont {Je}},\ }\href@noop
  {} {\bibfield  {journal} {\bibinfo  {journal} {IEEE Transactions on Circuits
  and Systems I: Regular Papers}\ }\textbf {\bibinfo {volume} {60}},\ \bibinfo
  {pages} {867} (\bibinfo {year} {2013})}\BibitemShut {NoStop}%
\bibitem [{\citenamefont {Chabalko}, \citenamefont {Shahmohammadi},\ and\
  \citenamefont {Sample}(2017)}]{Chabalko2017}%
  \BibitemOpen
  \bibfield  {author} {\bibinfo {author} {\bibfnamefont {M.~J.}\ \bibnamefont
  {Chabalko}}, \bibinfo {author} {\bibfnamefont {M.}~\bibnamefont
  {Shahmohammadi}}, \ and\ \bibinfo {author} {\bibfnamefont {A.~P.}\
  \bibnamefont {Sample}},\ }\href@noop {} {\bibfield  {journal} {\bibinfo
  {journal} {PloS one}\ }\textbf {\bibinfo {volume} {12}},\ \bibinfo {pages}
  {e0169045} (\bibinfo {year} {2017})}\BibitemShut {NoStop}%
\bibitem [{\citenamefont {Sasatani}\ \emph {et~al.}(2017)\citenamefont
  {Sasatani}, \citenamefont {Chabalko}, \citenamefont {Kawahara},\ and\
  \citenamefont {Sample}}]{Sastani2017}%
  \BibitemOpen
  \bibfield  {author} {\bibinfo {author} {\bibfnamefont {T.}~\bibnamefont
  {Sasatani}}, \bibinfo {author} {\bibfnamefont {M.~J.}\ \bibnamefont
  {Chabalko}}, \bibinfo {author} {\bibfnamefont {Y.}~\bibnamefont {Kawahara}},
  \ and\ \bibinfo {author} {\bibfnamefont {A.~P.}\ \bibnamefont {Sample}},\
  }\href@noop {} {\bibfield  {journal} {\bibinfo  {journal} {IEEE Antennas and
  Wireless Propagation Letters}\ }\textbf {\bibinfo {volume} {16}},\ \bibinfo
  {pages} {2746} (\bibinfo {year} {2017})}\BibitemShut {NoStop}%
\bibitem [{\citenamefont {Mei}\ \emph {et~al.}(2017)\citenamefont {Mei},
  \citenamefont {Thackston}, \citenamefont {Bercich}, \citenamefont
  {Jefferys},\ and\ \citenamefont {Irazoqui}}]{mei2017cavity}%
  \BibitemOpen
  \bibfield  {author} {\bibinfo {author} {\bibfnamefont {H.}~\bibnamefont
  {Mei}}, \bibinfo {author} {\bibfnamefont {K.~A.}\ \bibnamefont {Thackston}},
  \bibinfo {author} {\bibfnamefont {R.~A.}\ \bibnamefont {Bercich}}, \bibinfo
  {author} {\bibfnamefont {J.~G.}\ \bibnamefont {Jefferys}}, \ and\ \bibinfo
  {author} {\bibfnamefont {P.~P.}\ \bibnamefont {Irazoqui}},\ }\href@noop {}
  {\bibfield  {journal} {\bibinfo  {journal} {IEEE Transactions on Biomedical
  Engineering}\ }\textbf {\bibinfo {volume} {64}},\ \bibinfo {pages} {775}
  (\bibinfo {year} {2017})}\BibitemShut {NoStop}%
\bibitem [{\citenamefont {Kurs}\ \emph {et~al.}(2007)\citenamefont {Kurs},
  \citenamefont {Karalis}, \citenamefont {Moffatt}, \citenamefont
  {Joannopoulos}, \citenamefont {Fisher},\ and\ \citenamefont
  {Solja{\v{c}}i{\'c}}}]{Kurs2007}%
  \BibitemOpen
  \bibfield  {author} {\bibinfo {author} {\bibfnamefont {A.}~\bibnamefont
  {Kurs}}, \bibinfo {author} {\bibfnamefont {A.}~\bibnamefont {Karalis}},
  \bibinfo {author} {\bibfnamefont {R.}~\bibnamefont {Moffatt}}, \bibinfo
  {author} {\bibfnamefont {J.~D.}\ \bibnamefont {Joannopoulos}}, \bibinfo
  {author} {\bibfnamefont {P.}~\bibnamefont {Fisher}}, \ and\ \bibinfo {author}
  {\bibfnamefont {M.}~\bibnamefont {Solja{\v{c}}i{\'c}}},\ }\href@noop {}
  {\bibfield  {journal} {\bibinfo  {journal} {science}\ }\textbf {\bibinfo
  {volume} {317}},\ \bibinfo {pages} {83} (\bibinfo {year} {2007})}\BibitemShut
  {NoStop}%
\bibitem [{\citenamefont {Karalis}, \citenamefont {Joannopoulos},\ and\
  \citenamefont {Solja{\v{c}}i{\'c}}(2008)}]{Karalis2008}%
  \BibitemOpen
  \bibfield  {author} {\bibinfo {author} {\bibfnamefont {A.}~\bibnamefont
  {Karalis}}, \bibinfo {author} {\bibfnamefont {J.~D.}\ \bibnamefont
  {Joannopoulos}}, \ and\ \bibinfo {author} {\bibfnamefont {M.}~\bibnamefont
  {Solja{\v{c}}i{\'c}}},\ }\href@noop {} {\bibfield  {journal} {\bibinfo
  {journal} {Annals of Physics}\ }\textbf {\bibinfo {volume} {323}},\ \bibinfo
  {pages} {34} (\bibinfo {year} {2008})}\BibitemShut {NoStop}%
\bibitem [{\citenamefont {Song}\ \emph {et~al.}(2016)\citenamefont {Song},
  \citenamefont {Iorsh}, \citenamefont {Kapitanova}, \citenamefont
  {Nenasheva},\ and\ \citenamefont {Belov}}]{Song2016}%
  \BibitemOpen
  \bibfield  {author} {\bibinfo {author} {\bibfnamefont {M.}~\bibnamefont
  {Song}}, \bibinfo {author} {\bibfnamefont {I.}~\bibnamefont {Iorsh}},
  \bibinfo {author} {\bibfnamefont {P.}~\bibnamefont {Kapitanova}}, \bibinfo
  {author} {\bibfnamefont {E.}~\bibnamefont {Nenasheva}}, \ and\ \bibinfo
  {author} {\bibfnamefont {P.}~\bibnamefont {Belov}},\ }\href@noop {}
  {\bibfield  {journal} {\bibinfo  {journal} {Applied Physics Letters}\
  }\textbf {\bibinfo {volume} {108}},\ \bibinfo {pages} {023902} (\bibinfo
  {year} {2016})}\BibitemShut {NoStop}%
\bibitem [{\citenamefont {Song}, \citenamefont {Belov},\ and\ \citenamefont
  {Kapitanova}(2016)}]{Song2016Collosal}%
  \BibitemOpen
  \bibfield  {author} {\bibinfo {author} {\bibfnamefont {M.}~\bibnamefont
  {Song}}, \bibinfo {author} {\bibfnamefont {P.}~\bibnamefont {Belov}}, \ and\
  \bibinfo {author} {\bibfnamefont {P.}~\bibnamefont {Kapitanova}},\
  }\href@noop {} {\bibfield  {journal} {\bibinfo  {journal} {Applied Physics
  Letters}\ }\textbf {\bibinfo {volume} {109}},\ \bibinfo {pages} {223902}
  (\bibinfo {year} {2016})}\BibitemShut {NoStop}%
\bibitem [{\citenamefont {Elnaggar}(2017)}]{Elnaggar2017JAP}%
  \BibitemOpen
  \bibfield  {author} {\bibinfo {author} {\bibfnamefont {S.~Y.}\ \bibnamefont
  {Elnaggar}},\ }\href@noop {} {\bibfield  {journal} {\bibinfo  {journal}
  {Journal of Applied Physics}\ }\textbf {\bibinfo {volume} {121}},\ \bibinfo
  {pages} {064903} (\bibinfo {year} {2017})}\BibitemShut {NoStop}%
\bibitem [{\citenamefont {Hamam}\ \emph {et~al.}(2009)\citenamefont {Hamam},
  \citenamefont {Karalis}, \citenamefont {Joannopoulos},\ and\ \citenamefont
  {Solja{\v{c}}i{\'c}}}]{Hamam2009}%
  \BibitemOpen
  \bibfield  {author} {\bibinfo {author} {\bibfnamefont {R.~E.}\ \bibnamefont
  {Hamam}}, \bibinfo {author} {\bibfnamefont {A.}~\bibnamefont {Karalis}},
  \bibinfo {author} {\bibfnamefont {J.}~\bibnamefont {Joannopoulos}}, \ and\
  \bibinfo {author} {\bibfnamefont {M.}~\bibnamefont {Solja{\v{c}}i{\'c}}},\
  }\href@noop {} {\bibfield  {journal} {\bibinfo  {journal} {Annals of
  Physics}\ }\textbf {\bibinfo {volume} {324}},\ \bibinfo {pages} {1783}
  (\bibinfo {year} {2009})}\BibitemShut {NoStop}%
\bibitem [{\citenamefont {Zhang}\ \emph {et~al.}(2011)\citenamefont {Zhang},
  \citenamefont {Hackworth}, \citenamefont {Fu}, \citenamefont {Li},
  \citenamefont {Mao},\ and\ \citenamefont {Sun}}]{Zhang2012}%
  \BibitemOpen
  \bibfield  {author} {\bibinfo {author} {\bibfnamefont {F.}~\bibnamefont
  {Zhang}}, \bibinfo {author} {\bibfnamefont {S.~A.}\ \bibnamefont
  {Hackworth}}, \bibinfo {author} {\bibfnamefont {W.}~\bibnamefont {Fu}},
  \bibinfo {author} {\bibfnamefont {C.}~\bibnamefont {Li}}, \bibinfo {author}
  {\bibfnamefont {Z.}~\bibnamefont {Mao}}, \ and\ \bibinfo {author}
  {\bibfnamefont {M.}~\bibnamefont {Sun}},\ }\href {\doibase
  10.1109/TMAG.2010.2087010} {\bibfield  {journal} {\bibinfo  {journal} {IEEE
  Transactions on Magnetics}\ }\textbf {\bibinfo {volume} {47}},\ \bibinfo
  {pages} {1478} (\bibinfo {year} {2011})}\BibitemShut {NoStop}%
\bibitem [{\citenamefont {Elnaggar}, \citenamefont {Tervo},\ and\ \citenamefont
  {Mattar}(2014{\natexlab{a}})}]{Elnaggar2014coupled}%
  \BibitemOpen
  \bibfield  {author} {\bibinfo {author} {\bibfnamefont {S.~Y.}\ \bibnamefont
  {Elnaggar}}, \bibinfo {author} {\bibfnamefont {R.}~\bibnamefont {Tervo}}, \
  and\ \bibinfo {author} {\bibfnamefont {S.~M.}\ \bibnamefont {Mattar}},\
  }\href@noop {} {\bibfield  {journal} {\bibinfo  {journal} {Journal of
  Magnetic Resonance}\ }\textbf {\bibinfo {volume} {238}},\ \bibinfo {pages}
  {1} (\bibinfo {year} {2014}{\natexlab{a}})}\BibitemShut {NoStop}%
\bibitem [{\citenamefont {Elnaggar}, \citenamefont {Tervo},\ and\ \citenamefont
  {Mattar}(2014{\natexlab{b}})}]{Elnaggar2014Quality}%
  \BibitemOpen
  \bibfield  {author} {\bibinfo {author} {\bibfnamefont {S.~Y.}\ \bibnamefont
  {Elnaggar}}, \bibinfo {author} {\bibfnamefont {R.}~\bibnamefont {Tervo}}, \
  and\ \bibinfo {author} {\bibfnamefont {S.~M.}\ \bibnamefont {Mattar}},\
  }\href@noop {} {\bibfield  {journal} {\bibinfo  {journal} {Journal of
  Magnetic Resonance}\ }\textbf {\bibinfo {volume} {242}},\ \bibinfo {pages}
  {57} (\bibinfo {year} {2014}{\natexlab{b}})}\BibitemShut {NoStop}%
\bibitem [{\citenamefont {Mattar}\ and\ \citenamefont
  {Elnaggar}(2017)}]{AMR2017}%
  \BibitemOpen
  \bibfield  {author} {\bibinfo {author} {\bibfnamefont {S.~M.}\ \bibnamefont
  {Mattar}}\ and\ \bibinfo {author} {\bibfnamefont {S.~Y.}\ \bibnamefont
  {Elnaggar}},\ }\href@noop {} {\bibfield  {journal} {\bibinfo  {journal}
  {Applied Magnetic Resonance}\ }\textbf {\bibinfo {volume} {48}},\ \bibinfo
  {pages} {1205} (\bibinfo {year} {2017})}\BibitemShut {NoStop}%
\bibitem [{\citenamefont {Elnaggar}, \citenamefont {Saha},\ and\ \citenamefont
  {Antar}(2019)}]{samehwptmeas}%
  \BibitemOpen
  \bibfield  {author} {\bibinfo {author} {\bibfnamefont {S.~Y.}\ \bibnamefont
  {Elnaggar}}, \bibinfo {author} {\bibfnamefont {C.}~\bibnamefont {Saha}}, \
  and\ \bibinfo {author} {\bibfnamefont {Y.~M.}\ \bibnamefont {Antar}},\ }\href
  {\doibase 10.1063/1.5129280} {\bibfield  {journal} {\bibinfo  {journal}
  {Journal of Applied Physics}\ }\textbf {\bibinfo {volume} {126}},\ \bibinfo
  {pages} {244902} (\bibinfo {year} {2019})},\ \Eprint
  {http://arxiv.org/abs/https://doi.org/10.1063/1.5129280}
  {https://doi.org/10.1063/1.5129280} \BibitemShut {NoStop}%
\bibitem [{\citenamefont {Saha}, \citenamefont {Elnaggar},\ and\ \citenamefont
  {Antar}(2018)}]{wptc1}%
  \BibitemOpen
  \bibfield  {author} {\bibinfo {author} {\bibfnamefont {C.}~\bibnamefont
  {Saha}}, \bibinfo {author} {\bibfnamefont {S.}~\bibnamefont {Elnaggar}}, \
  and\ \bibinfo {author} {\bibfnamefont {Y.}~\bibnamefont {Antar}},\ }in\
  \href@noop {} {\emph {\bibinfo {booktitle} {Wireless Power Transfer (WPTC),
  2018 IEEE International Conference on}}}\ (\bibinfo {organization} {IEEE},\
  \bibinfo {year} {2018})\BibitemShut {NoStop}%
\bibitem [{\citenamefont {Elnaggar}, \citenamefont {Saha},\ and\ \citenamefont
  {Antar}(2018)}]{wptc2}%
  \BibitemOpen
  \bibfield  {author} {\bibinfo {author} {\bibfnamefont {S.}~\bibnamefont
  {Elnaggar}}, \bibinfo {author} {\bibfnamefont {C.}~\bibnamefont {Saha}}, \
  and\ \bibinfo {author} {\bibfnamefont {Y.}~\bibnamefont {Antar}},\ }in\
  \href@noop {} {\emph {\bibinfo {booktitle} {Wireless Power Transfer (WPTC),
  2018 IEEE International Conference on}}}\ (\bibinfo {organization} {IEEE},\
  \bibinfo {year} {2018})\BibitemShut {NoStop}%
\bibitem [{\citenamefont {Roberts}(1947)}]{robertsdielectric}%
  \BibitemOpen
  \bibfield  {author} {\bibinfo {author} {\bibfnamefont {S.}~\bibnamefont
  {Roberts}},\ }\href@noop {} {\bibfield  {journal} {\bibinfo  {journal}
  {Physical Review}\ }\textbf {\bibinfo {volume} {71}},\ \bibinfo {pages} {890}
  (\bibinfo {year} {1947})}\BibitemShut {NoStop}%
\bibitem [{\citenamefont {Yim}\ \emph {et~al.}(2015)\citenamefont {Yim},
  \citenamefont {Yong}, \citenamefont {Lee}, \citenamefont {Lee}, \citenamefont
  {Nahm}, \citenamefont {Yoo}, \citenamefont {Lee}, \citenamefont {Hwang},\
  and\ \citenamefont {Han}}]{yimnature}%
  \BibitemOpen
  \bibfield  {author} {\bibinfo {author} {\bibfnamefont {K.}~\bibnamefont
  {Yim}}, \bibinfo {author} {\bibfnamefont {Y.}~\bibnamefont {Yong}}, \bibinfo
  {author} {\bibfnamefont {J.}~\bibnamefont {Lee}}, \bibinfo {author}
  {\bibfnamefont {K.}~\bibnamefont {Lee}}, \bibinfo {author} {\bibfnamefont
  {H.-H.}\ \bibnamefont {Nahm}}, \bibinfo {author} {\bibfnamefont
  {J.}~\bibnamefont {Yoo}}, \bibinfo {author} {\bibfnamefont {C.}~\bibnamefont
  {Lee}}, \bibinfo {author} {\bibfnamefont {C.~S.}\ \bibnamefont {Hwang}}, \
  and\ \bibinfo {author} {\bibfnamefont {S.}~\bibnamefont {Han}},\ }\href@noop
  {} {\bibfield  {journal} {\bibinfo  {journal} {NPG Asia Materials}\ }\textbf
  {\bibinfo {volume} {7}},\ \bibinfo {pages} {e190} (\bibinfo {year}
  {2015})}\BibitemShut {NoStop}%
\bibitem [{\citenamefont {Si3N4}\ \emph {et~al.}(1999)\citenamefont {Si3N4}
  \emph {et~al.}}]{singh}%
  \BibitemOpen
  \bibfield  {author} {\bibinfo {author} {\bibfnamefont {S.}~\bibnamefont
  {Si3N4}} \emph {et~al.},\ }\href@noop {} {\bibfield  {journal} {\bibinfo
  {journal} {Electrochemical Society Interface}\ ,\ \bibinfo {pages} {27}}
  (\bibinfo {year} {1999})}\BibitemShut {NoStop}%
\bibitem [{\citenamefont {Shende}\ \emph {et~al.}(2001)\citenamefont {Shende},
  \citenamefont {Krueger}, \citenamefont {Rossetti},\ and\ \citenamefont
  {Lombardo}}]{shende2001strontium}%
  \BibitemOpen
  \bibfield  {author} {\bibinfo {author} {\bibfnamefont {R.~V.}\ \bibnamefont
  {Shende}}, \bibinfo {author} {\bibfnamefont {D.~S.}\ \bibnamefont {Krueger}},
  \bibinfo {author} {\bibfnamefont {G.~A.}\ \bibnamefont {Rossetti}}, \ and\
  \bibinfo {author} {\bibfnamefont {S.~J.}\ \bibnamefont {Lombardo}},\
  }\href@noop {} {\bibfield  {journal} {\bibinfo  {journal} {Journal of the
  American Ceramic Society}\ }\textbf {\bibinfo {volume} {84}},\ \bibinfo
  {pages} {1648} (\bibinfo {year} {2001})}\BibitemShut {NoStop}%
\bibitem [{\citenamefont {Elnaggar}, \citenamefont {Tervo},\ and\ \citenamefont
  {Mattar}(2015{\natexlab{a}})}]{Elnaggar2015ECMT}%
  \BibitemOpen
  \bibfield  {author} {\bibinfo {author} {\bibfnamefont {S.~Y.}\ \bibnamefont
  {Elnaggar}}, \bibinfo {author} {\bibfnamefont {R.~J.}\ \bibnamefont {Tervo}},
  \ and\ \bibinfo {author} {\bibfnamefont {S.~M.}\ \bibnamefont {Mattar}},\
  }\href@noop {} {\bibfield  {journal} {\bibinfo  {journal} {IEEE Transactions
  on Microwave Theory and Techniques}\ }\textbf {\bibinfo {volume} {63}},\
  \bibinfo {pages} {2115} (\bibinfo {year} {2015}{\natexlab{a}})}\BibitemShut
  {NoStop}%
\bibitem [{\citenamefont {Elnaggar}, \citenamefont {Tervo},\ and\ \citenamefont
  {Mattar}(2015{\natexlab{b}})}]{Elnaggar2015JAP}%
  \BibitemOpen
  \bibfield  {author} {\bibinfo {author} {\bibfnamefont {S.~Y.}\ \bibnamefont
  {Elnaggar}}, \bibinfo {author} {\bibfnamefont {R.~J.}\ \bibnamefont {Tervo}},
  \ and\ \bibinfo {author} {\bibfnamefont {S.~M.}\ \bibnamefont {Mattar}},\
  }\href@noop {} {\bibfield  {journal} {\bibinfo  {journal} {Journal of Applied
  Physics}\ }\textbf {\bibinfo {volume} {118}},\ \bibinfo {pages} {194901}
  (\bibinfo {year} {2015}{\natexlab{b}})}\BibitemShut {NoStop}%
\bibitem [{\citenamefont {Pozar}(2011)}]{Pozar05}%
  \BibitemOpen
  \bibfield  {author} {\bibinfo {author} {\bibfnamefont {D.~M.}\ \bibnamefont
  {Pozar}},\ }\href {http://amazon.com/o/ASIN/0470631554/} {\emph {\bibinfo
  {title} {Microwave Engineering}}},\ \bibinfo {edition} {4th}\ ed.\ (\bibinfo
  {publisher} {Wiley},\ \bibinfo {year} {2011})\BibitemShut {NoStop}%
\bibitem [{\citenamefont {Elnaggar}, \citenamefont {Tervo},\ and\ \citenamefont
  {Mattar}(2015{\natexlab{c}})}]{Elnaggar2015Image}%
  \BibitemOpen
  \bibfield  {author} {\bibinfo {author} {\bibfnamefont {S.~Y.}\ \bibnamefont
  {Elnaggar}}, \bibinfo {author} {\bibfnamefont {R.~J.}\ \bibnamefont {Tervo}},
  \ and\ \bibinfo {author} {\bibfnamefont {S.~M.}\ \bibnamefont {Mattar}},\
  }\href@noop {} {\bibfield  {journal} {\bibinfo  {journal} {IEEE Transactions
  on Microwave Theory and Techniques}\ }\textbf {\bibinfo {volume} {63}},\
  \bibinfo {pages} {2124} (\bibinfo {year} {2015}{\natexlab{c}})}\BibitemShut
  {NoStop}%
\bibitem [{\citenamefont {Popovi\'{c}}, \citenamefont {Manolatou},\ and\
  \citenamefont {Watts}(2006)}]{Popovic06}%
  \BibitemOpen
  \bibfield  {author} {\bibinfo {author} {\bibfnamefont {M.~A.}\ \bibnamefont
  {Popovi\'{c}}}, \bibinfo {author} {\bibfnamefont {C.}~\bibnamefont
  {Manolatou}}, \ and\ \bibinfo {author} {\bibfnamefont {M.~R.}\ \bibnamefont
  {Watts}},\ }\href {\doibase 10.1364/OE.14.001208} {\bibfield  {journal}
  {\bibinfo  {journal} {Opt. Express}\ }\textbf {\bibinfo {volume} {14}},\
  \bibinfo {pages} {1208} (\bibinfo {year} {2006})}\BibitemShut {NoStop}%
\end{thebibliography}%
\smallskip
\end{document}